\documentclass[conference]{IEEEtran}
\IEEEoverridecommandlockouts

\usepackage{booktabs}
\usepackage{multirow}
\usepackage{subcaption}
\usepackage[dvipsnames]{xcolor}
\usepackage{url}
\usepackage{graphicx}
\usepackage{hyperref}
\usepackage{stix}
\usepackage{tcolorbox}

\AtBeginDocument{%
  }

\begin{document}

\title{Social Media Reactions to Open Source Promotions:\\ AI-Powered GitHub Projects on Hacker News}

\author{Prachnachai Meakpaiboonwattana\textsuperscript{\dag}, Warittha Tarntong\textsuperscript{\dag}, Thai Mekratanavorakul\textsuperscript{\dag}, Chaiyong Ragkhitwetsagul\textsuperscript{\dag},\\ Pattaraporn Sangaroonsilp\textsuperscript{\dag}, Raula Kula\textsuperscript{\ddag}, Morakot Choetkiertikul\textsuperscript{\dag}, Kenichi Matsumoto\textsuperscript{*}, Thanwadee Sunetnanta\textsuperscript{\dag}
\\
\textsuperscript{\dag}\textit{Faculty of Information and Communication Technology, Mahidol University, Thailand}\\
\textsuperscript{\ddag}\textit{Graduate School of Information Science and Technology, The University of Osaka, Japan}\\
\textsuperscript{*}\textit{Graduate School of Science and Technology, NAIST, Japan}}

\maketitle

\begin{abstract}
Social media platforms have become more influential than traditional news sources, shaping public discourse and accelerating the spread of information. With the rapid advancement of artificial intelligence (AI), open-source software (OSS) projects can leverage these platforms to gain visibility and attract contributors. In this study, we investigate the relationship between Hacker News, a social news site focused on computer science and entrepreneurship, and the extent to which it influences developer activity on the promoted GitHub AI projects.
We analyzed 2,195 Hacker News (HN) stories and their corresponding comments over a two-year period. Our findings reveal that at least 19\% of AI developers promoted their GitHub projects on Hacker News, often receiving positive engagement from the community.
By tracking activity on the associated 1,814 GitHub repositories after they were shared on Hacker News, we observed a significant increase in forks, stars, and contributors. These results suggest that Hacker News serves as a viable platform for AI-powered OSS projects, with the potential to gain attention, foster community engagement, and accelerate software development.
\end{abstract}


\maketitle

\section{Introduction}
\label{sec:introduction}
Large Language Models (LLMs), a type of generative AI, have captivated the world since the launch of ChatGPT on November 30, 2022. Researchers and industries are now exploring ways to utilize this potent language model for their work. This has led to an exponential increase in applications developed to support or integrate AIs~\cite{Bergmann2024}. GitHub reported a staggering 65,000 public generative AI projects created in 2023, marking a 284\% year-over-year growth~\cite{Octoverse2023}. Furthermore, contributions to generative AI projects surged by 59\% in 2024, while the total number of AI projects rose by 98\%~\cite{Octoverse2024}.

Open-source developers employ various methods to promote their projects publicly. This study centers on Hacker News (HN), a technology and entrepreneurship-focused social website~\cite{HackerNewsFAQ,Combinatora}. HN is often used by AI project creators to gain attention and stimulate community discussions~\cite{AIRisingNow}.

\begin{figure*}
    \centering
    \includegraphics[width=1.8\columnwidth]{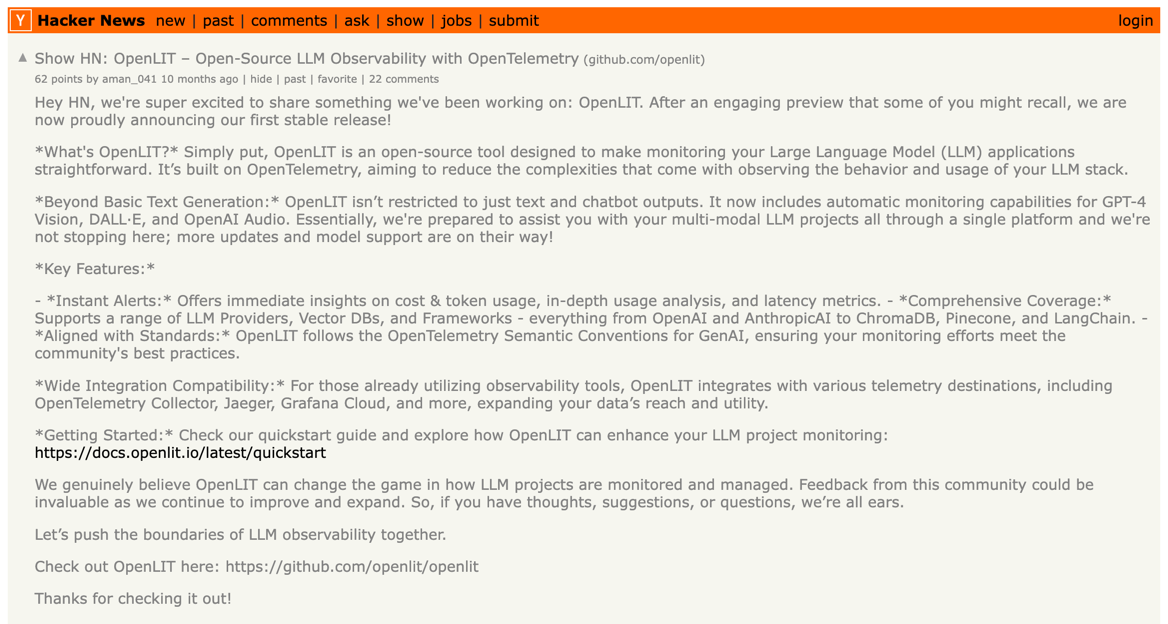}
    \caption{Hacker News story \#40167461 to promote their OpenLIT project}
    \label{fig:example}
\end{figure*}

A motivating example is presented in Figure~\ref{fig:example}. It illustrates a Hacker News story advertising the open-source LLM project, OpenLIT\footnote{\url{https://news.ycombinator.com/item?id=40167461}}. The poster follows the convention of using ``Show HN'' in the title to present intriguing work to the community. The story provides a summary of the project, its key features, and a link to OpenLIT's GitHub repository (\url{https://github.com/openlit/openlit}). Posted on April 26, 2024, it sparked some discussion (22 comments).
The HN community displayed mixed reactions towards this project in the comments. Some users expressed enthusiasm and requested additional details, while others questioned its practicality. The poster or other users occasionally addressed these concerns. In this paper, we refer to such stories as ``Hacker News GitHub AI stories (HN GH-AI stories).''

Hacker News has been a subject of interest in software engineering research, with studies exploring different aspects: Stoddard et al.\cite{stoddard2015popularity} investigated the relationship between popularity and quality of online articles on Hacker News and Reddit. Aniche et al.\cite{Aniche2018} analyzed the characteristics and community dynamics of these platforms, emphasizing Hacker News's distinctive, non-personalized community vote ranking system. Additionally, Barik et al.~\cite{Barik2015} compared automatic information extraction from Hacker News to grounded theory research, underscoring its efficiency and reliability as a source for empirical software engineering studies.

In this study, we mined and analyzed Hacker News stories pertaining to AI projects to investigate the social reactions of the HN community towards these projects and their impact on project activities. We aim to answer the following research questions.

\begin{itemize}
    \item \textbf{RQ1: What is the spread of GitHub AI project stories on Hacker News?}
We aim to understand the number of HN GH-AI stories on Hacker News and their trends over time by analyzing 2,195 HN stories and 4,476 comments related to open-source AI projects on GitHub over 2 years (From May 8, 2022, to May 9, 2024).
    
\item \textbf{RQ2: What are the social reactions to HN GH-AI stories?}
We aim to gain insights into the attitudes of the posters of HN stories and the user comments associated with AI projects. Using sentiment analysis, we investigated the overall reaction of HN GH-AI stories and comments.

\item \textbf{RQ3: What are the changes of activities in GitHub AI projects after being mentioned in Hacker News?} 
We study the GitHub activity metrics to understand the impact of promotion on the project.
 
\end{itemize}

Understanding these research questions (RQs) provides insights into how GitHub AI projects are shared and discussed on HN.
We made 8 observations and insights that also aligned with the research questions.
First, GitHub AI stories were widespread on HN, particularly following the release of ChatGPT. These stories were often posted close to the creation of their respective repositories, sometimes by the contributors themselves, highlighting a strong correlation between notable AI project releases and their visibility on HN.
Second, the overall sentiment towards GitHub AI projects on HN was neutral, yet there were more positive stories and comments than negative ones. Notably, some particularly strong reactions indicated that users had tried and liked the projects, suggesting that positive user engagement often translates to online enthusiasm.
Finally, projects with positive sentiment exhibited increased activity, such as spikes in stars, forks, and contributors, shortly after being posted on HN. 


These findings collectively illustrate the role of HN in promoting and shaping public discourse around AI projects, influencing their growth and adoption through community interactions and sentiment.
The contribution of this work is as follows:

\begin{enumerate}
    \item Our study is the first to reveal how AI developers use Hacker News to advertise their projects and the reception from the community.
    \item Our study provides insights based on the repository information. We observed increased activity on the GitHub projects across several metrics, i.e., commits, pull requests, stars, forks, and contributors, after they were posted on Hacker News. 
    \item The dataset collected from Hacker News and GitHub are made publicly available~\cite{replication}.
\end{enumerate}


\begin{figure*}
    \centering
    \includegraphics[width=1\linewidth]{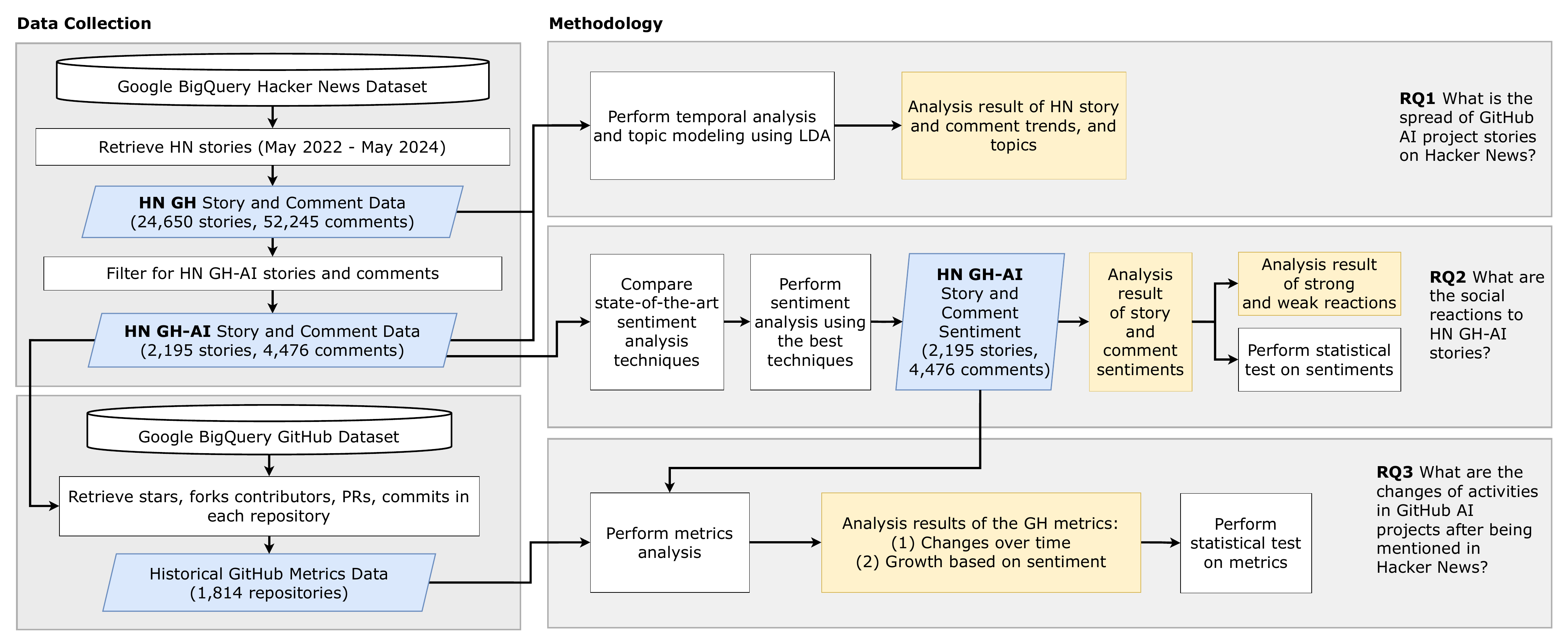}
    \caption{Overview of Data Preparation and Methodology}
    \label{fig:method-outline}
\end{figure*}

\section{Hacker News For Self-Promotion of OSS Projects}
\label{sec:HackerNews}

Referring back to Figure~\ref{fig:example}, we can see that the story contains a title that often reflects the headline of the linked content (``\textit{Show HN: OpenLIT - Open-Source LLM Observability with OpenTelemetry}''). The title can be followed by a URL that links to external content such as articles, blog posts, or GitHub repositories. In this case, it shows a link to the GitHub account of OpenLIT's owner (``github.com/openlit''). Below the story (not shown in the figure) is a comment section where HN users discuss the story. Each comment contains the metadata, including submission time, vote count, and user information. There can be comments that respond to stories and also other comments.

The Hacker News platform organizes content through ``\textit{stories},'' which can be linked to external content or text-based discussions. Each story appears on the site's front page based on a ranking algorithm that considers factors such as vote count, submission time, and comment activity. 

We consider using Hacker News in our study for several reasons. First, it curates significant AI developments through community voting, where important projects gain genuine attention from the community rather than being promoted by recommendation algorithms, as seen on typical social media platforms. Second, its comment system offers valuable qualitative insights into how developers and technology enthusiasts perceive and respond to AI advancements. Third, it features GitHub links that help track AI adoption in open-source projects.


\section{Study Design}
\label{sec:dataprep}
\subsection{Data Preparation}
\paragraph{Collecting Hacker News Data}
We collected HN stories and comments using Google Cloud's BigQuery~\cite{BigQueryEnterpriseData} as part of it being available as a public dataset~\cite{HackerNewsMarketplace} (see Figure \ref{fig:method-outline}).  Past research on modern news aggregators, including HN, suggested BigQuery usage to overcome the rate limits for the retrieval process~\cite{anicheHowModernNews2018}. The criteria for the story query are that the story must neither be ``dead'' nor ``deleted'', which were indicated by their associated \textit{dead} and \textit{deleted} flag columns, and that it must contain an associated URL, story title, and author. The story retrieval spanned 2 years from May 8, 2022, 6 months prior to ChatGPT release~\cite{RiseOfChatGPT}, to May 9, 2024, a year and a half after ChatGPT release, encompassing a significant period in AI development.




Furthermore, we filtered only for stories containing GitHub repository URLs (i.e., HN GH stories) to focus specifically on software projects. We applied regular expressions on each of the story's URL. Multiple filters were used in this case, with the first pattern checking for \texttt{github.com} in the parsed URL. Additional regular expression patterns were used to distinguish between repository URLs and other URLs, such as issue, pull request, commit, and branch URLs.

Based on the filtered HN GH stories, we filtered for HN GH-AI stories using another set of regular expressions. To filter only for relevant stories, we established a set of keywords based on (1) frequent AI and LLM-specific terms from existing research (e.g., \texttt{chatgpt}, \texttt{llm}, \texttt{nlp})~\cite{minaee2024largelanguagemodelssurvey}, and (2) additional keywords related to recent LLM models (e.g., \texttt{claude}, \texttt{gemini}).  The final list of keywords used for the HN story collection is presented in Table \ref{tab:keywords}. Stories whose titles corresponded to at least one of the keywords were collected. Note that each keyword must be matched with word boundaries, such as spaces or underscores, before and after itself to prevent potential false matches. For instance, the story title \texttt{BRAIN} should not match the keyword \texttt{AI} because it does not have any word boundaries before and after the keyword. Additionally, certain keywords that may overlap with non-AI stories were manually excluded, such as \texttt{palm} and \texttt{transformer}, due to them not being AI-exclusive keywords, hence the chance of them falsely matching with non-AI story titles.

\begin{table}[bt]
    \centering
    \caption{List of AI keywords used for filtering Hacker News stories}
    \begin{tabular}{l}
        \toprule
        Keywords \\
        \midrule
        \texttt{artificial intelligence}, \texttt{ai}, \\ 
        \texttt{natural language processing}, \texttt{nlp}, \texttt{language model}, \\ 
        \texttt{llm}, \texttt{chatbot}, \texttt{chatgpt}, \texttt{openai}, \texttt{claude}, \texttt{llama}, \texttt{gemini}, \\
        \texttt{gemma}, \texttt{mistral}, \texttt{cohere} \\
        \bottomrule
        \end{tabular}%
    \label{tab:keywords}
\end{table}


Using the keywords matching mechanism, the other remaining HN stories containing GitHub URLs are considered not related to AI. 
One limitation of this approach is that some stories may not have been identified as HN GH-AI stories using our set of AI-related keywords. As a result, there could be additional AI-related stories that were not captured.
Nonetheless, we have established that HN GH-AI stories would be the minimum amount of GitHub AI stories on HN.
After filtering, we collected the comments using BigQuery. Using the similar criteria as the stories, we only retrieve non-empty comments associated with the collected stories and flagged as neither \textit{dead} nor \textit{deleted}.
Only first-level comments were retrieved because they directly addressed the HN GH-AI stories rather than responding to other comments.

The final Hacker News data collected for this study contains 24,650 total HN GH stories (stories containing GitHub URLs), among which contains 2,195 HN GH-AI stories and their associated 4,476 comments (as shown in Table~\ref{tab:dataset-stats}).


\begin{table}[bt]
    \centering
    \caption{The Hacker News and GitHub data used in this study}
    \begin{tabular}{p{7cm}r}
        \toprule
        \textbf{Hacker News dataset} & \textbf{\#} \\
        \midrule
        Hacker News GitHub stories (HN GH stories) & 24,650 \\
        Hacker News GitHub stories' comments (HN GH comments) & 52,245 \\
        \midrule
        \textbf{Filtered Hacker News dataset (RQ1 \& RQ2)} & \\
        \midrule
        Hacker News GitHub AI stories (HN GH-AI stories) & 2,195 \\
        Hacker News GitHub AI stories' comments (HN GH-AI comments) & 4,476 \\
        \midrule
        \textbf{Historical GitHub Metric dataset (RQ3)} & \\
        \midrule
        GitHub repositories with historical metrics data & 1,814 \\
        \hline
    \end{tabular}
    \small
    \label{tab:dataset-stats}
\end{table}

\paragraph{Constructing Hacker News Sentiment Ground Truth}

Since we could not find any existing HN sentiment ground truth dataset in the literature, we manually constructed a representative ground truth dataset of HN stories and comments for selecting the best sentiment analysis technique for our study (later discussed in Section~\ref{sec:methodology}). 385 stories were uniformly selected with a 95\% confidence interval. Within each story, comments were sampled using stratified sampling. This sampling resulted in 385 comments collected. 

For labeling, we employed two human investigators (two authors of this study who have four years of experience in software development).
To obtain human labels, two investigators established labeling criteria and independently labeled each story and comment.
Then, the two investigators compared their results.
The inter-rater reliability between the two investigators (Cohen’s Kappa coefficient~\cite{mchughInterraterReliabilityKappa2012}) was 0.626 for comments and 0.719 for stories, both of which indicate substantial agreement~\cite{Cohen-kappa-interpret-landis-koch-1977}.
Then, they engaged in detailed discussions to resolve disagreements and generate consensus labels.

\paragraph{Collecting Historical GitHub Metrics Data}
Using HN GH-AI stories, which include GitHub repository URLs, we retrieved historical GitHub metrics from BigQuery for each repository. The data collection period spans from May 2022, approximately six months before ChatGPT's release, to January 2025, which is about six months after our most recent HN GH-AI story retrieval (May 2024). Each repository is expected to have 33 monthly entries, with each entry containing metric changes compared to the previous month. The collected metrics include the following:

\begin{itemize}
    \item \textbf{Stars}: Number of new stars.
    \item \textbf{Forks}: Number of new forks.
    \item \textbf{Commits}: Number of new commits.
    \item \textbf{Contributors}: The project's new contributors.
    \item \textbf{Pull Requests (PRs)}: Number of new pull requests (both opened and merged).
\end{itemize}

To ensure data integrity, we removed duplicate repository URLs, as some GitHub repositories appeared in multiple HN GH-AI stories. This prevents redundant entries and ensures that each repository is uniquely represented in the dataset.

\subsection{Research Methodology to Answer RQs}
\label{sec:methodology}

The overview of our research methodology is shown in Figure \ref{fig:method-outline}. It outlines the steps taken to answer each research question based on the collected Hacker News and historical GitHub metrics data. The white boxes represent the processes used to gather the data in each step, and the blue boxes represent the collected data. Lastly, the yellow boxes represent the analysis performed on the collected data.

\paragraph{Answering RQ1}



We used the Hacker News dataset to answer RQ1.
To see the relationship of whether the HN GH-AI posters were the GitHub repository creators themselves, we matched the GitHub repository contributor account names with the HN story posters' accounts. This includes direct matches between GitHub accounts, as well as matches from social media handles, GitHub profile URLs, and emails that may be presented on the HN GH-AI story posters' \textit{About} section. The matching was performed automatically, and the researchers did not try to identify the HN posters. We only performed a few manual random inspections to make sure that our matching worked correctly. 



\textbf{Temporal Trend Analysis:}
To analyze the spread of AI project stories through Hacker News, we performed a temporal trend analysis and a topic modeling analysis. These analyses aim to discover the quantitative growth patterns and the qualitative evolution of discourse around AI projects.

Using the collected HN GH and HN GH-AI stories, a series of temporal analyses was conducted to gain a better understanding of the AI project trends on Hacker News. We count the frequency of the HN GH-AI stories in each 14-day bin period, as well as compare them with the overall HN GH projects. This enables us to identify whether the proportion of GitHub AI projects has been increasing over the years relative to the overall GH projects on Hacker News.
Furthermore, to understand the rough distribution of keyword matches among the HN GH-AI stories, we identified the number of stories containing each of the keywords in Table \ref{tab:keywords} and their occurrences over time.



\textbf{Topic Modeling Analysis:}
We applied topic modeling to the story titles. 
Among various topic modeling approaches, we chose Latent Dirichlet Allocation (LDA) for its interpretability and effectiveness in handling short texts like titles \cite{blei2003latent}. Unlike other methods, such as Non-negative Matrix Factorization (NMF) or Latent Semantic Analysis (LSA), LDA provides probabilistic topic assignments and has shown superior performance in capturing semantic relationships in similar studies of technical discussions \cite{hanWhatProgrammersDiscuss2020}.
A previous study~\cite{Silva2021_topicModellingSE} shows that 
LDA-based techniques are the most commonly used approaches
along with a manual topic naming based on frequent keywords.

The preprocessing pipeline for the story titles used the NLTK library~\cite{bird2009natural} and consisted of several steps: (1) text normalization (conversion to lowercase), (2) removal of special characters and digits, (3) tokenization, (4) stopword removal, (5) lemmatization and stemming, (6) and removal of empty titles.
 To focus on crucially distinct terms, we removed \texttt{show hn} and \texttt{ai} terms on the story titles from our topic modeling analysis. 
As mentioned in the Section~\ref{sec:introduction}, the ``Show HN'' title convention is generally used to showcase projects on Hacker News. Therefore, many HN GH-AI stories are expected to contain the convention in their title.

A TF-IDF matrix was then constructed to identify the top 100 terms based on the TF-IDF scores from the HN titles. To optimize processing efficiency while maintaining analytical integrity, we applied UMAP for dimensionality reduction \cite{UMAPUniformManifold}. This reduced representation served as input for subsequent analyses.
Then, the optimal number of topics was determined through an evaluation of topic coherence scores. We evaluated coherence scores for different numbers of topics (ranging from 2 to 20) and selected the number that maximized the NPMI coherence values of the resulting topics, which are expected to yield as clear distinction as possible between each topic. The NPMI coherence measure was chosen as it combines both semantic similarity and co-occurrence statistics, providing a more robust evaluation of topic quality \cite{roder2015exploring}.

\paragraph{Answering RQ2}
We used the Filtered Hacker News dataset in this step.
To understand the reception of AI projects on Hacker News, we developed an approach for sentiment analysis to categorize stories and their comments as ``positive towards AI'' (+1), ``negative towards AI'' (-1), or ``neutral'' (0). 

We only analyzed the first-level comments directed specifically at the HN GH-AI stories rather than the comments that respond to other comments. This is because direct responses to the original story would more accurately reflect sentiment toward the AI project, whereas a comment such as a simple response like ``I agree'' to another comment would not provide meaningful insight into the project's reception. 

HN stories and comments are text-only and cannot contain images or complex media. Potential noises in HN comments include simple HTML entities (e.g., \texttt{\&quot;} for quotes) and basic tags, which we decoded before inputting them through LLMs for sentiment analysis. Other text preprocessing, such as stemming or lemmatization, was not applied to HN stories and comments as we experimented with fine-tuning pre-trained transformers and LLMs, which required contextual understanding to be effective. 

\textbf{Finding the Best Sentiment Analysis Technique for HN:}
We investigated many state-of-the-art sentiment analysis techniques in the literature for this task. In order to find the best-performing technique for our dataset, we compared them using our manually labeled sentiment ground truth following the suggestions of the work of Lin et al. \cite{Lin2018}, where they warned that researchers should not assume the sentiment analysis tool provides a reliable result off-the-shelf. 

We compared three state-of-the-art transformer models for sentiment analysis, including Twitter-roBERTa~\cite{camacho-collados_tweetnlp_2022}, RoBERTa~\cite{Liu2019_roBERTa}, and BERT~\cite{Devlin2018-bert}, and one LLM, OpenAI's GPT-4o mini~\cite{OpenAIPlatformModels}, using our ground truth dataset. We applied all models to the HN stories and comments in the ground truth and calculated their weighted F1 scores. The three state-of-the-art sentiment analysis techniques were fine-tuned using 5-fold cross-validation to optimize their hyperparameters. For GPT-4o mini, we used few-shot prompting techniques to guide the LLM's answer\footnote{The prompts that we used can be found in our replication package~\cite{replication}} and tested it on the ground truth dataset.

The results, presented in Table~\ref{tab:comment_sentiment_f1_result}, show that GPT-4o mini outperformed the other models for both the story and comment sentiment. For stories, GPT-4o mini offers the highest weighted F1 score (0.762), followed by Twitter-RoBERTa (0.733), RoBERTa (0.724), and BERT (0.683). For comments, GPT-4o mini offers the highest weighted F1 score similar to the stories (0.763), followed by Twitter-RoBERTa (0.709), RoBERTa (0.681), and BERT (0.599). Based on these findings, we selected GPT-4o mini for our sentiment analysis.
Our finding complements the result reported by Zhang et al.~\cite{Zhang2024}, who show that LLMs yield strong performance in simple sentiment analysis tasks and even outperform fine-tuned small language models when data is scarce and by Zhang et al.~\cite{Zhang2025}, who show that prompted LLMs has state-of-the-art performance when data is limited and unbalanced.
To see the sentimental landscape of the HN community, each sentiment (negative, neutral, and positive) of HN GH-AI stories and comments was plotted over time using an area stack chart.

\begin{table}[bt]
    \centering
    \caption{Comparison of sentiment analysis performance (F1-score) on HN comments and stories}
    \begin{tabular}{lrr}
        \toprule
        Model & Stories & Comments \\
        \midrule
         BERT & 0.683 & 0.599\\
         RoBERTa & 0.724 & 0.681 \\
         Twitter-RoBERTa & 0.733 & 0.709 \\
         GPT-4o mini & \textbf{0.762} & \textbf{0.763} \\
         \bottomrule
    \end{tabular}
    \label{tab:comment_sentiment_f1_result}
\end{table}

\textbf{Analyzing the HN Social Reaction to GitHub AI stories:}
To gauge whether a comment provides a strong reaction or not, we defined relationships between the sentiment of the comments based on the sentiment of the stories and categorized them into either \textbf{strong reaction (+2)} or \textbf{weak reaction (+1)}. As shown in Table~\ref{tab:reactions}, if the comment's sentiment is either positive or negative, disregarding the story's sentiment, this creates a strong reaction. If the story and the comment are both neutral, then the comment is considered a weak reaction. Please note that the reaction scores emphasize the strength of the reactions and do not differentiate between positive or negative sentiments. 

\begin{table}[bt]
    \centering
    \caption{Classification of HN reactions}
    \begin{tabular}{lll}
    \toprule
    Story sentiment & Comment sentiment & Reaction \\
    \midrule
    Positive & Positive/Negative & Strong reaction (+2) \\
    Positive & Neutral & Weak reaction (+1) \\
    \midrule
    Negative & Positive/Negative & Strong reaction (+2) \\
    Negative & Neutral & Weak reaction (+1) \\
    \midrule
    Neutral & Positive/Negative & Strong reaction (+2)\\
    Neutral & Neutral & Weak reaction (+1) \\
    \bottomrule
    \end{tabular}
    \label{tab:reactions}
\end{table}

\paragraph{Answering RQ3}
\textbf{Calculating the GitHub Metrics Changes:}
Using the Historical GitHub Metric dataset, we studied the trends of changes for each of the five GitHub metrics (contributors, forks, stars, commits, and pull requests).
To understand repository evolution over time, we calculated the number of monthly changes in each of the five metrics by subtracting the value of the previous month. For example, if a GitHub repository A had 100 stars in February 2024 and 80 stars in January 2024, the number of star change is 20. We applied this to all the collected GitHub metrics over 2 years.

For each month in the Historical GitHub Metric dataset, we identified and removed outliers using the interquartile range (IQR) method. Unlike statistical dispersion measures such as the z-score, which assumes a normal distribution \cite{13517DetectionOutliers}, GitHub metrics can exhibit volatility and skewness from month to month. Therefore, IQR is more appropriate for outlier detection in this study \cite{uptonUnderstandingStatistics1996}.
Additionally, we excluded entries where all metric values were zero before conducting the analysis. A zero value may indicate that the repository has not yet been created, and removing such entries helps highlight meaningful metric changes in visualizations.

\textbf{Analyzing the Potential Impact of Hacker News Exposure:}
We investigated the potential impact of HN exposure by comparing repository metric changes after their submission on Hacker News. 
Based on the monthly-change data of the five metrics, we chose only the data that occurred after the GitHub project was posted on HN. This spans over a 5-month period, including 1 month, 2 months, 3 months, 4 months, and 5 months after their HN submission date.
Then, we grouped the metrics based on the comment sentiment that the HN stories received.
This methodology enables us to compare the changes in GitHub projects' metrics by the social engagement on HN, uncovering patterns and trends that may be affected by being exposed to the HN community.

\section{Results}
\label{sec:results}
This section discusses our observations of the findings, as well as the answers to the research questions.


\subsubsection{Analysis of the Hacker News Posters and Posting Time}
\begin{figure}
    \centering
    \includegraphics[width=1\linewidth]{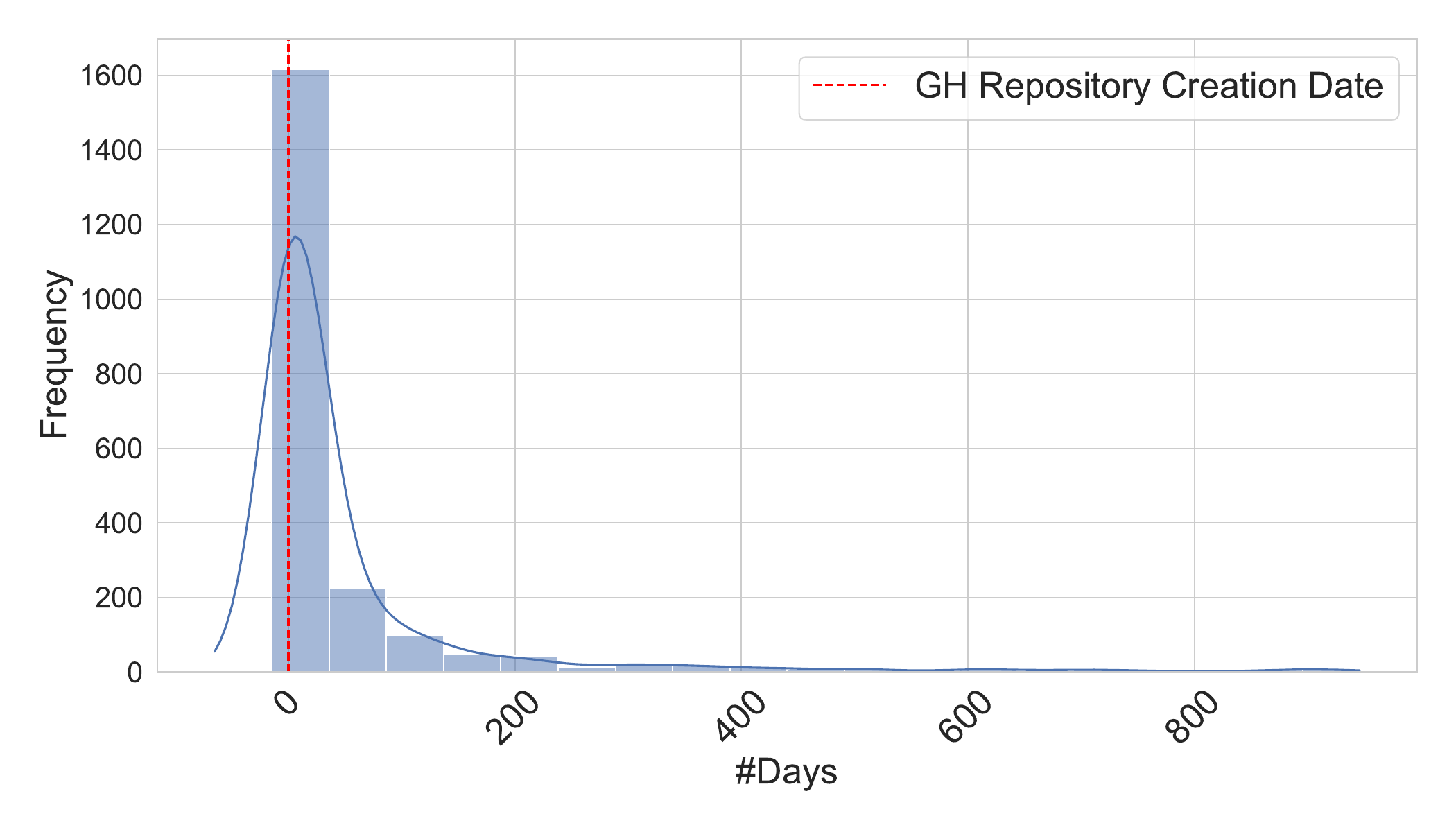}
    \caption{Difference (days) between HN submission and GH repo creation}
    \label{fig:hn-gh-submission-histogram}
\end{figure}

\begin{figure}
    \centering
    \includegraphics[width=1\linewidth]{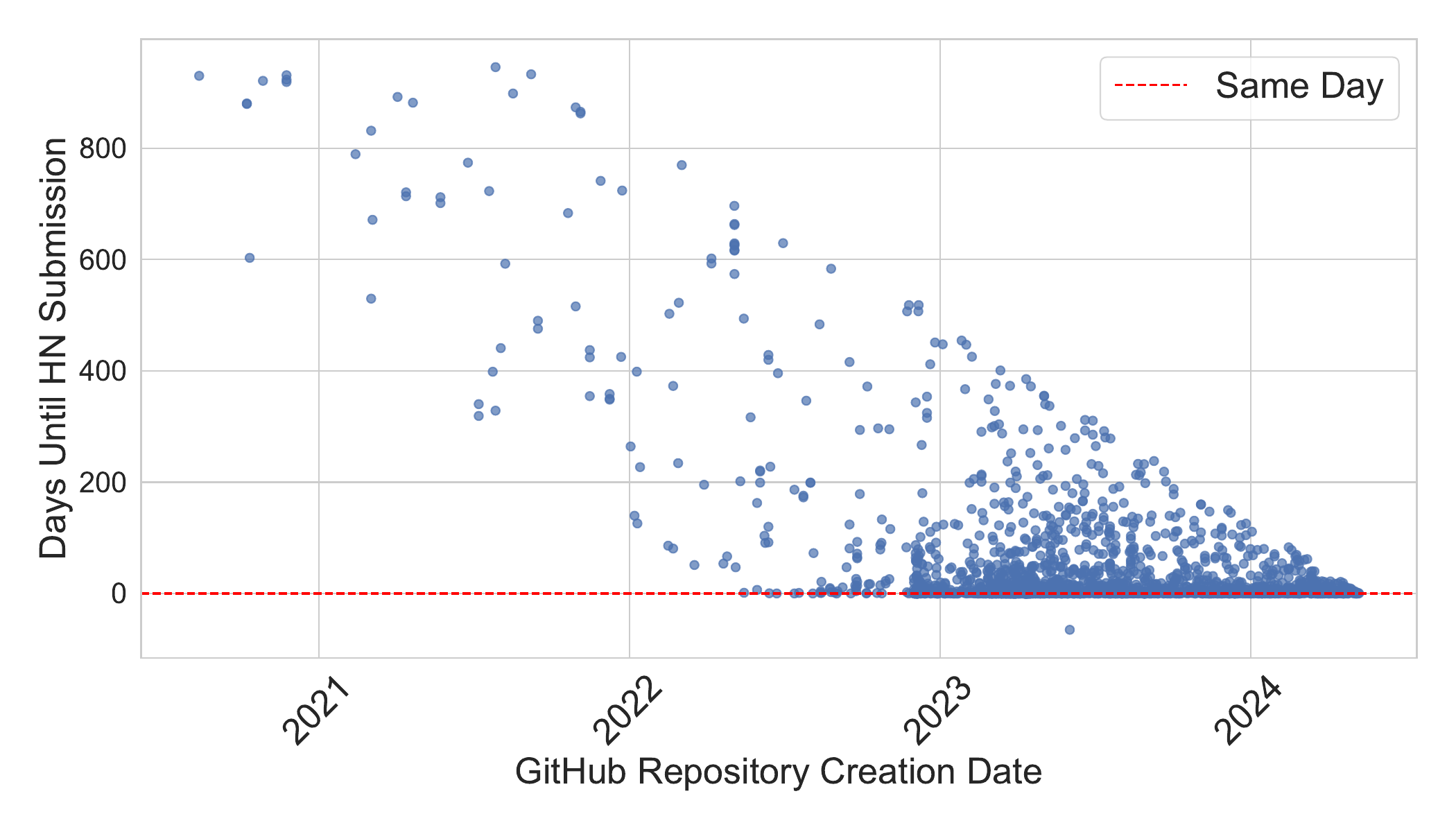}
    \caption{Time to HN submission vs.~GH repo creation date}
    \label{fig:hn-gh-submission-scatterplot}
\end{figure}

We found that most HN users tend to submit HN GH-AI stories within one week of their repository creation, as shown in Figure~\ref{fig:hn-gh-submission-histogram}. More specifically, Figure~\ref{fig:hn-gh-submission-scatterplot} illustrates that more recent GitHub AI projects, particularly those created between 2023 and 2024, are submitted to HN on gradually fewer days (from 0--600 days to 0--200 days) after their creation compared to projects created before 2023, which exhibit a broader range before Hacker News submission.

\begin{tcolorbox}
\textit{Observation 1: Majority of GitHub AI projects submitted on Hacker News close to the time of their creation on GitHub, implying there is no delay between creation and promotion.}
\end{tcolorbox}


Furthermore, investigation into the matching between HN GH-AI story posters and the associated GitHub repository owner reveals that a minimum of 471 out of all HN GH-AI posters (19\%) were the GitHub repository owner themselves. This provides additional evidence that a substantial amount of GitHub project owners currently see HN as an important medium to promote their projects and potentially gain an audience. Other reasons could be to gather early feedback from IT-enthusiasts demographics to be able to pivot their work more easily in the early stage. Future investigations may be required to understand the underlying reasons for this relationship and why the project owners have decided to submit to HN.

\begin{tcolorbox}
\textit{Observation 2: At least 19\% of the posters of HN GH-AI stories were contributors to the GitHub project themselves, implying self-promotion.}
\end{tcolorbox} 

\begin{figure*}
    \centering
    \begin{subfigure}[b]{0.49\textwidth}
        \includegraphics[width=\linewidth]{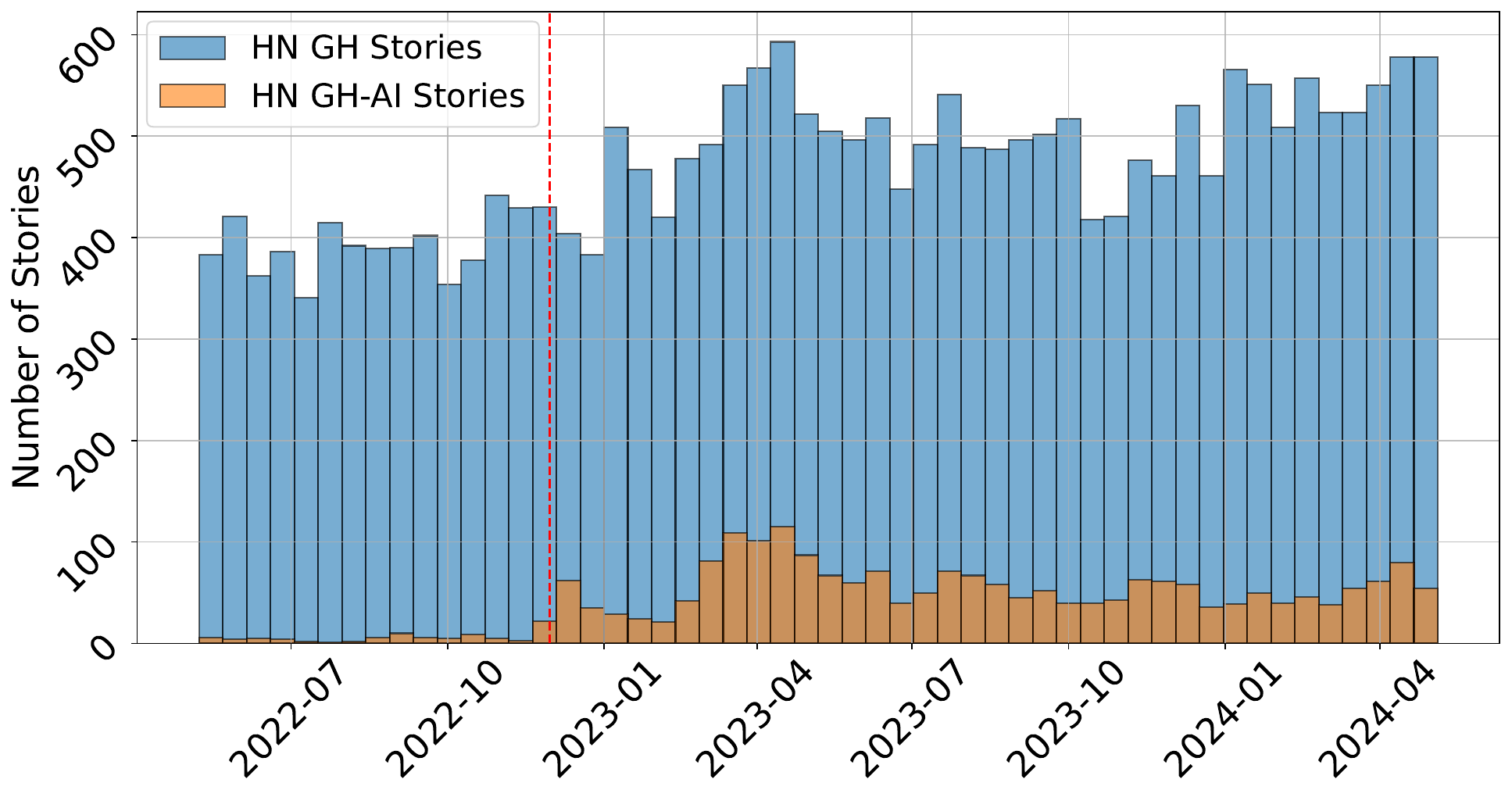}
        \caption{Trend of HN GH stories (bin=14 days)}
        \label{fig:hn-trends-gh}
    \end{subfigure}
    \begin{subfigure}[b]{0.49\textwidth}
        \includegraphics[width=\linewidth]{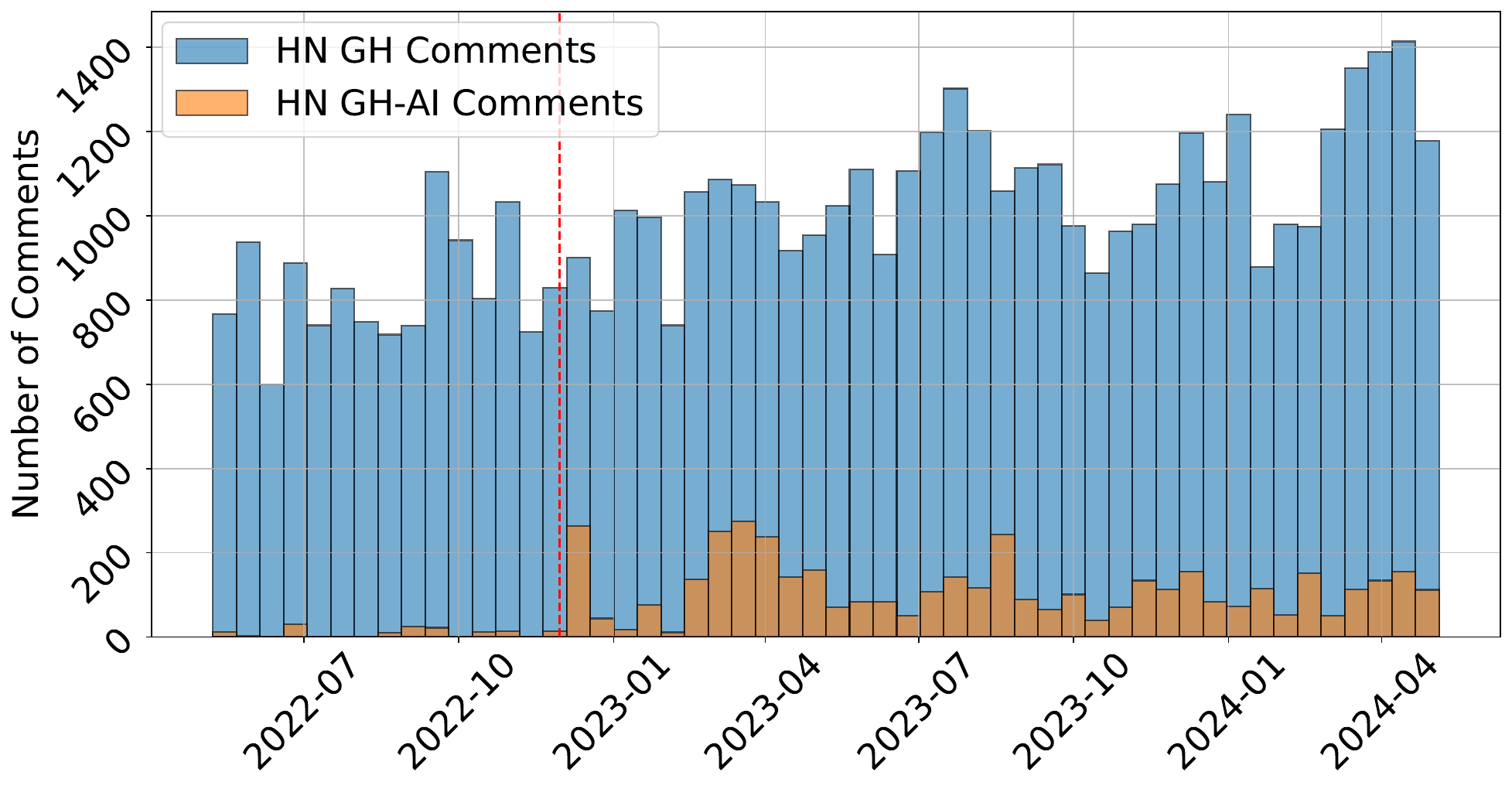}
        \caption{Trend of HN GH comments (bin=14 days)}
        \label{fig:hn-trends-gh-comments}
    \end{subfigure}
    
    \caption{Comparison histograms of HN GH and HN GH-AI Stories}
    \label{fig:rq1-hn-stories-comments-trends}
\end{figure*}




\subsubsection{Temporal Distribution of HN GH-AI Stories}
After analyzing 2,195 HN GH-AI stories and their 4,476 associated comments (previously shown in Table~\ref{tab:dataset-stats}) over two years, we observed some interesting trends, as follows.

Figure \ref{fig:hn-trends-gh} depicts the number of Hacker News' GitHub (i.e., HN GH) and Hacker News' GitHub-AI (HN GH-AI) stories over the years. The blue bars represent HN GH projects, which can include both AI and non-AI projects, while the yellow bars represent the HN GH-AI. We can observe that having GitHub URLs in HN stories was a common phenomenon even before ChatGPT's release date (marked by the dotted red line). However, the number of HN GH-AI has been rising after ChatGPT's release date. This suggests its significant impact on open-source AI development and advertisement in recent years. 



We observed the same trends in the HN comments. Figure \ref{fig:hn-trends-gh-comments} showed a notable increase in the number of comments on HN GH-AI projects (yellow) after the release of ChatGPT compared to the more stable trend of comments on HN stories containing other GitHub projects (blue).

\begin{tcolorbox}
Observation 3: HN GH-AI stories increased sharply after the ChatGPT release, confirming the impact of the LLMs on AI projects. 
\end{tcolorbox}


\begin{figure}
    \centering
    \includegraphics[width=1\linewidth]{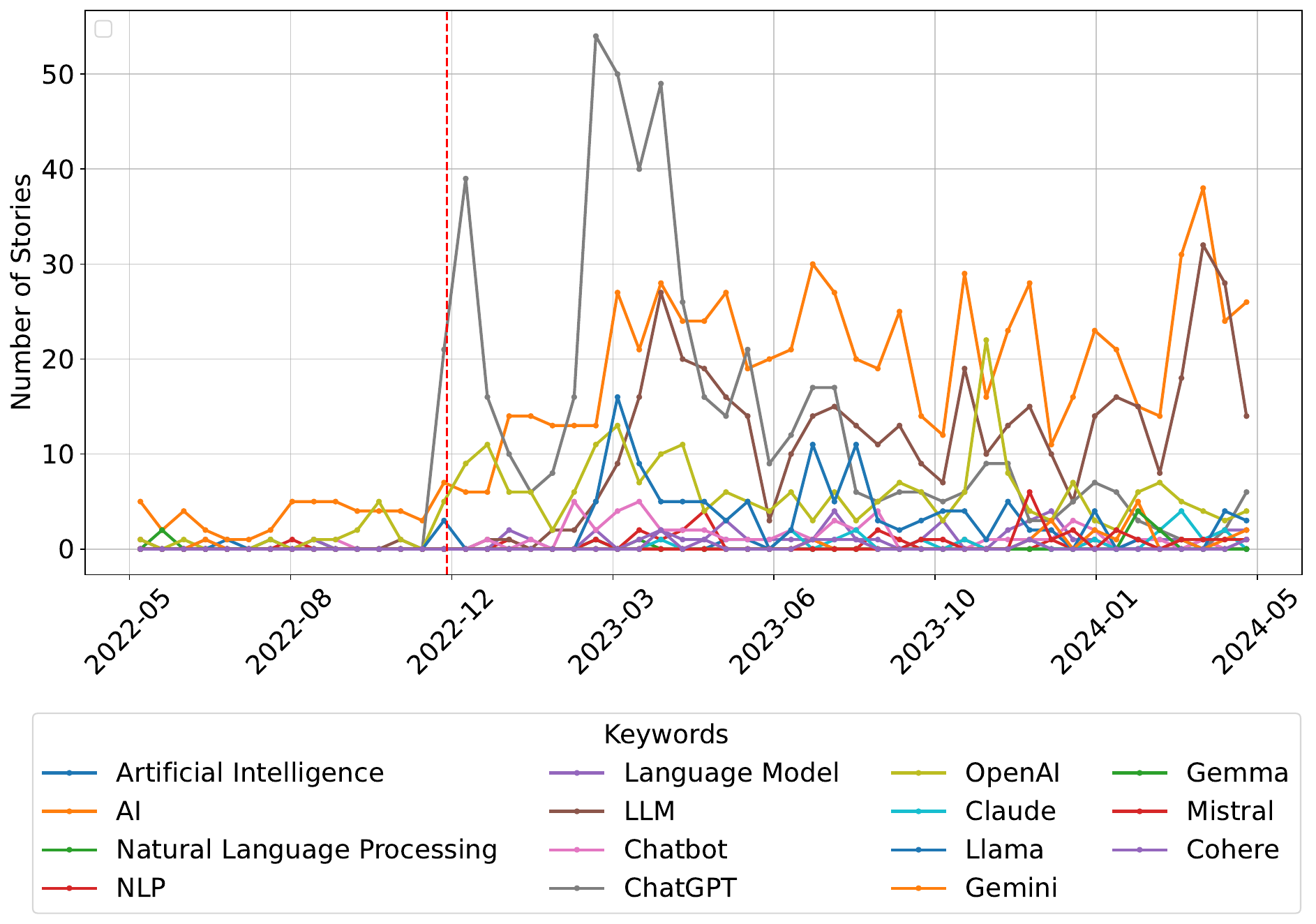}
    \caption{Keyword trends from HN GH-AI stories.}
    \label{fig:hn-trends-keywords}
\end{figure}

\subsubsection{Keyword trends and Topic Modeling of HN GH-AI Stories}
An analysis of the keyword distribution in the analyzed HN stories is illustrated in Figure \ref{fig:hn-trends-keywords}. It shows that \texttt{chatgpt} was a significant contributor to this increase, with distinct spikes in activity observed after both the ChatGPT release and the subsequent GPT-4 launch, indicating that many GitHub AI projects are related to OpenAI's products or platforms. Projects containing other LLM models are also mentioned on Hacker News, including \texttt{claude}, \texttt{gemini}, \texttt{llama}, \texttt{gemma}, \texttt{mistral}, and \texttt{cohere}.

We trained an LDA model with 8 topics using Scikit-learn, optimizing coherence score. The model used batch learning (learning\_decay=0.7), 30 iterations, and alpha/eta=`auto' for topic and word distribution adjustments. t-SNE visualization confirmed clear topic separability. As shown in Table~\ref{tab:topics-analysis}, the topics include AI projects using OpenAI's models, chatbots, plugins, automation, and LLM alternatives.

\begin{tcolorbox}
Observation 4: HN GH-AI stories cover 8 types of AI/LLM applications that range from LLM usage to model evaluations.
\end{tcolorbox}

\begin{table}
    \centering
    \caption{HN GH-AI Stories' Topic Terms Distribution}
    \begin{tabular}{lp{2.7cm}p{4.0cm}}
        \toprule
        \textbf{Topic} & \textbf{Top Terms} & \textbf{Interpretation} \\
        \midrule
        T1 & \texttt{openai, chatgpt, code, llm, function} & Projects using OpenAI's models, e.g., ChatGPT \\
        \midrule
        T2 & \texttt{open, sourc, chatbot, convers, llm} & Open-source chatbot projects and conversational modles \\
        \midrule
        T3 & \texttt{llm, chatgpt, openai, python, whisper} & LLM implementations, including OpenAI tools like Whisper and ChatGPT \\
        \midrule
        T4 & \texttt{chatgpt, plugin, command, use, line} & ChatGPT plugins, command-line usage, and integrations \\
        \midrule
        T5 & \texttt{list, ui, project, llm, chatgpt} & UI-driven projects and curated lists related to LLMs and ChatGPT \\
        \midrule
        T6 & \texttt{chatgpt, game, commit, automat, llama} & AI-powered automation using ChatGPT including games and commit reviews \\
        \midrule
        T7 & \texttt{languag, model, opensourc, llm, framework} & Open-source language models, frameworks, and LLM development \\
        \midrule
        T8 & \texttt{chatgpt, termin, llm, gpt, altern} & GPT alternatives, terminology, and model evaluations \\
        \bottomrule
    \end{tabular}
    \label{tab:topics-analysis}
\end{table}

\textbf{To answer RQ1, we found that the GitHub AI stories were widespread on Hacker News, especially right after the release of ChatGPT. They were mostly posted close to the time of the GitHub repository creation, and some of them were posted by the contributors of the repository themselves. The AI projects covered a wide range of AI applications, such as chatbot, ChatGPT plugins, UI-based tools, and automation tools.}

\subsubsection{Attitudes of HN Community Toward HN GH-AI Stories}
\begin{figure}[htbp]
    \centering
    \includegraphics[width=\linewidth]{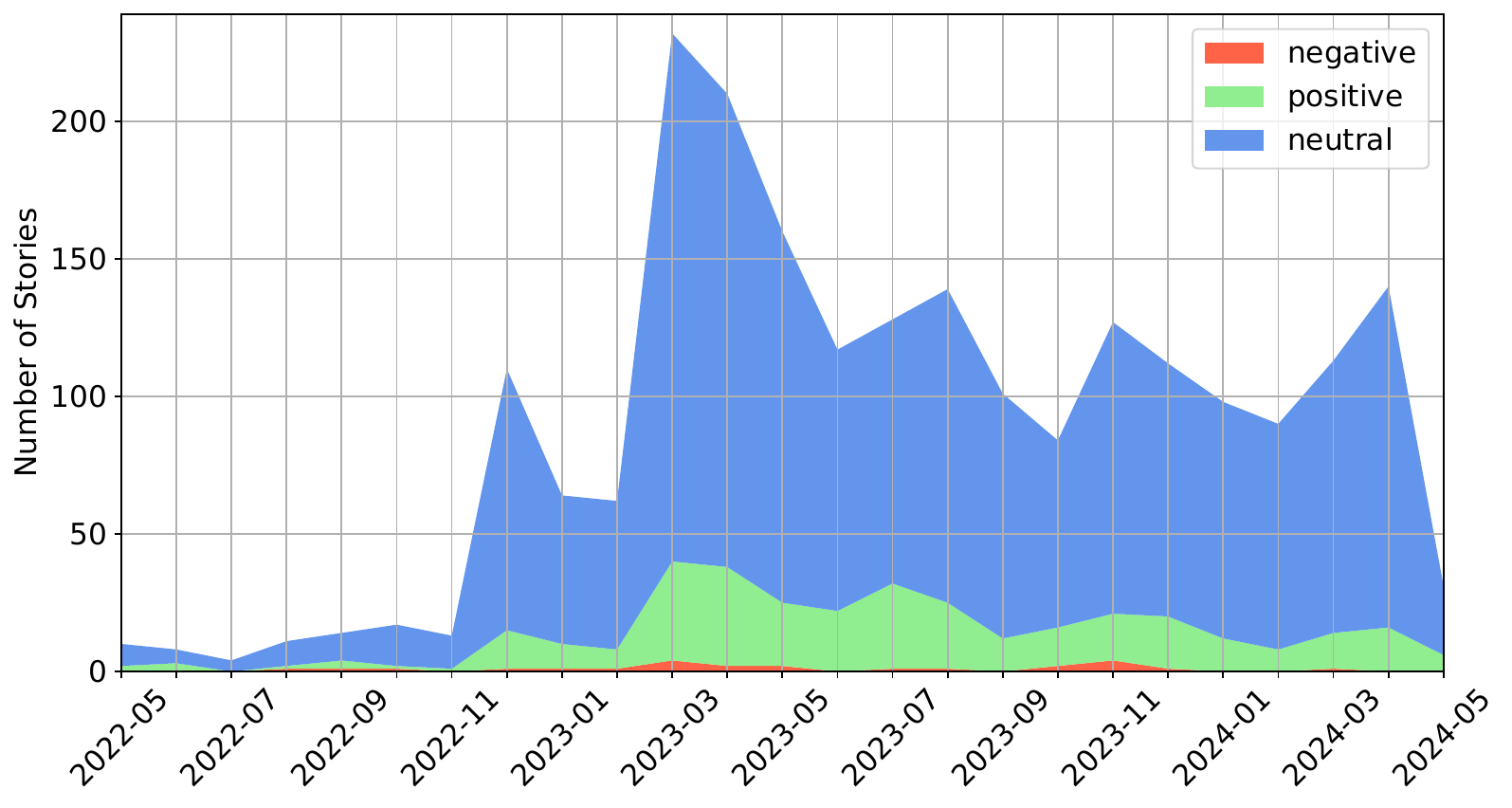}
    \caption{Sentiment of HN GH-AI stories}
    \label{fig:ai_story_area_stack}
\end{figure}

We found that the Hacker News community generally has positive attitudes toward stories related to AI-powered projects on GitHub.
As shown in Figure \ref{fig:ai_story_area_stack}, most HN GH-AI stories have neutral sentiment (blue), followed by positive (green) and negative (red). This is because the HN GH-AI stories are often announcements (with ``show hn'' keywords), which typically do not contain any attitude toward AI or the projects. 

\begin{figure}[htbp]
    \centering
    \includegraphics[width=\linewidth]{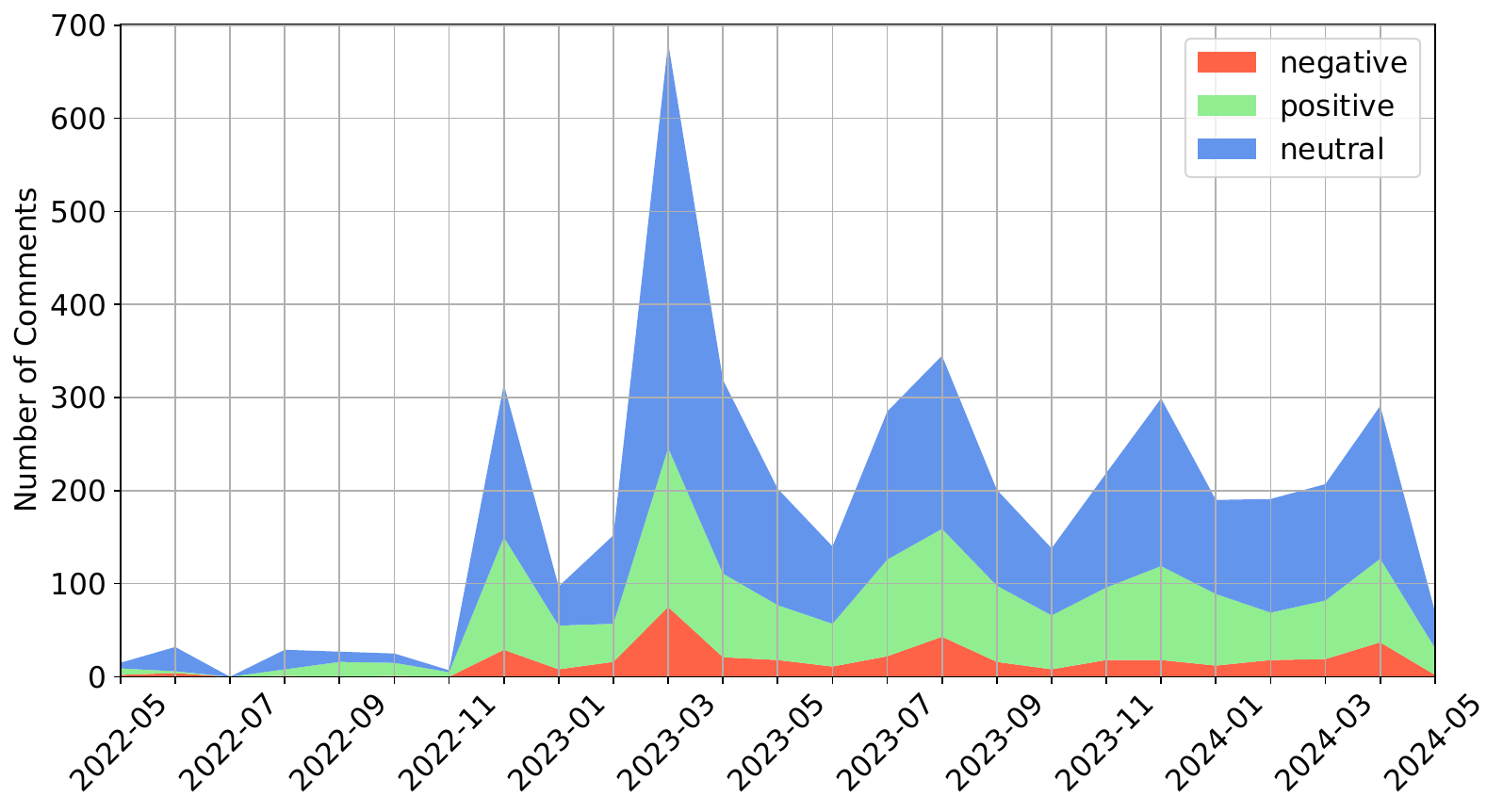}
    \caption{Sentiment of HN GH-AI comments}
    \label{fig:ai_comment_area_stack}
\end{figure}

The sentiment results of comments in HN GH-AI stories better reflect the attitudes of the HN community toward the announcements of AI projects than the story sentiment. This is because the comments are created by HN users who are not the owners of the GitHub projects. It shows how the community thinks about such projects being posted on Hacker News.
As shown in Figure~\ref{fig:ai_comment_area_stack}, disregarding the neutral sentiment, we also see more positive comments in HN GH-AI stories than negative ones. This observation applies throughout the whole analyzed period of two years. 

To support our finding, we performed a statistical test using non-parametric Pearson's chi-square ($X^2$) test between the sentiment of the collected HN GH-AI stories and 22,455 HN stories that do not contain AI keywords (the control group). We found a statistically significant difference for story (p-value = $1.32\times10^-25$) and comment sentiments (p-value = $1.95\times10^-10$). Thus, these observed sentiments of HN GH-AI stories and comments did not happen by chance.

\begin{tcolorbox}
Observation 5: HN GH-AI stories received more positive feedback than negative feedback on Hacker News.
\end{tcolorbox}

\begin{figure}[htbp]
    \centering
    \includegraphics[width=\linewidth]{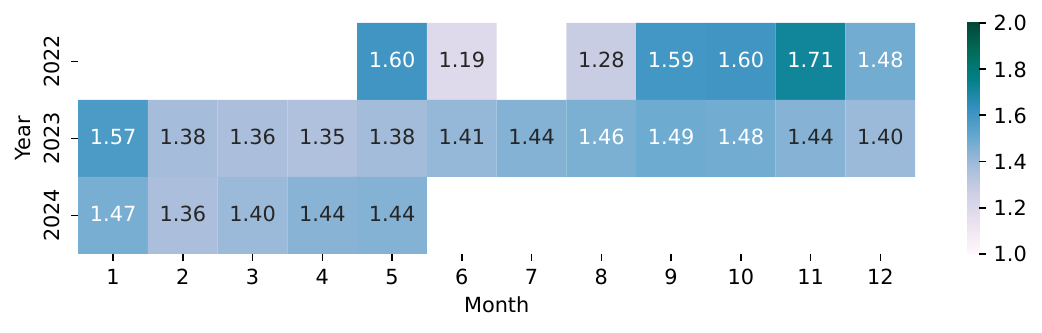}
    \caption{HN GH-AI with strong and weak reactions}
    \label{fig:ai_comment_heatmap}
\end{figure}

In addition, the results from our reaction analysis (see Table~\ref{tab:reactions}) show that the HN community had a few strong reactions (i.e., values close to 2.0) to HN GH-AI stories, as shown in Figure~\ref{fig:ai_comment_heatmap}.
We can spot two spikes in the reaction on May 2022 and around ChatGPT's release from September 2022 to January 2023.
Below are two examples of the HN GH-AI stories with strong reactions. These examples show that the commenters had already tried using the advertised projects and were happy with them.

Story\#31521467 \textit{``Show HN: Developer Friendly Natural Language Processing (github.com/winkjs)''} $\leftarrow$ Comment:~``\textit{\textbf{Huh, interesting. It gets POS tagging correct in a few edge cases I've used to test other frameworks. Bookmarked!}}'' 

Story\#33869231 \textit{``Show HN: Chrome extension to summarize blogs and articles using ChatGPT (github.com/clmnin)''} $\leftarrow$ Comment:~\textit{\textbf{``Great job, I was doing it manually before! It would be nice if it would be possible to continue chatting with chatGPT after the summary. I always find it interesting to probe chatGPT about the article after summarizing it''}}


\begin{tcolorbox}
Observation 6: The discussions on Hacker News regarding HN GH-AI stories show a few strong social reactions.
\end{tcolorbox}

\textbf{To answer RQ2, we found that, generally, the attitude toward GitHub AI projects on Hacker News is neutral. Nonetheless, there were more positive stories and comments than negative ones. A few of the strong reactions on Hacker News show that HN users already tried and were satisfied with the projects.}

\begin{figure*}
    \begin{subfigure}[b]{0.19\textwidth}
        \includegraphics[width=\linewidth]{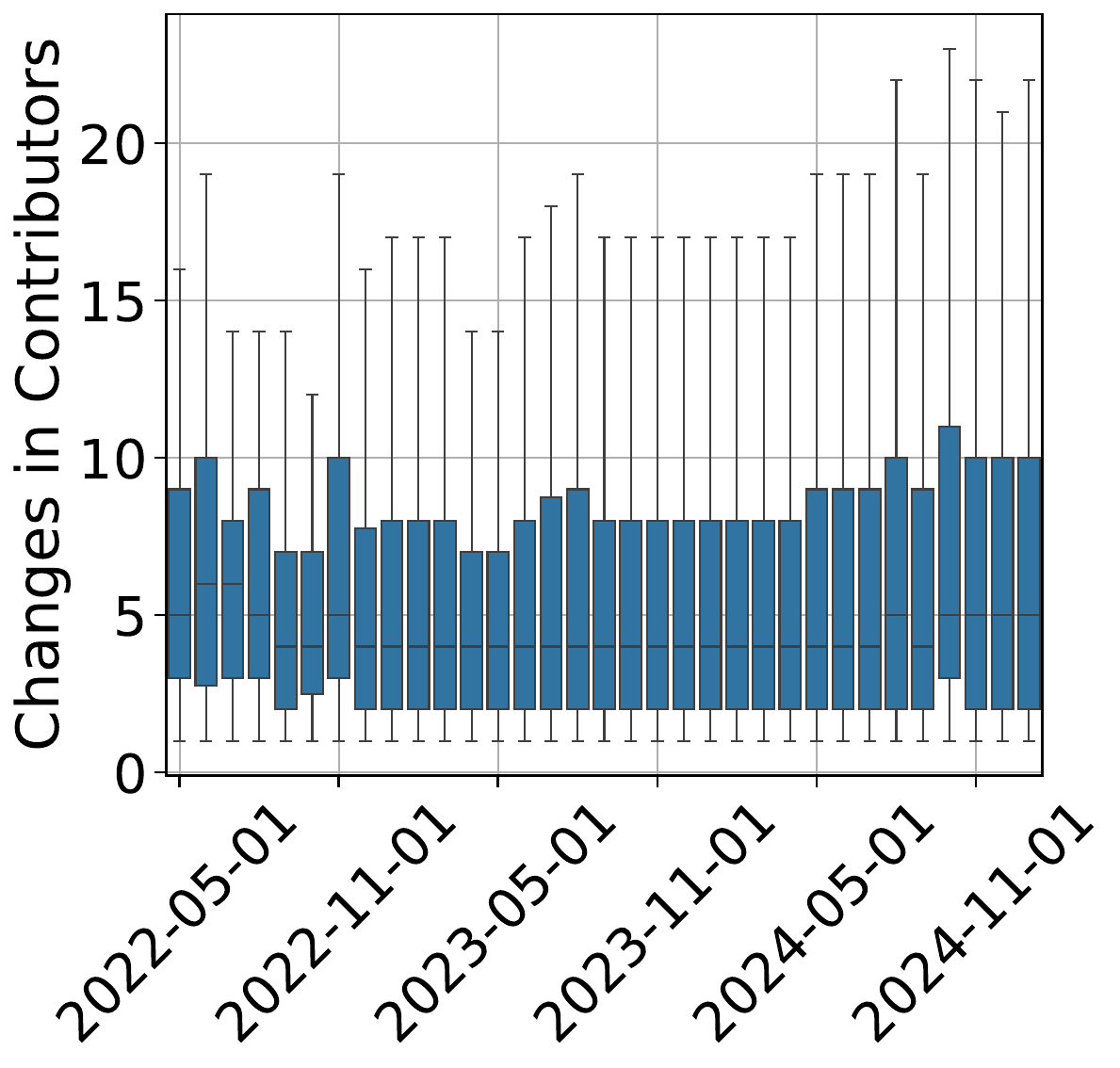}
        \caption{Contributors}
        \label{fig:rq3-boxplot-contributors}
    \end{subfigure}
    \hfill
    \begin{subfigure}[b]{0.19\textwidth}
        \includegraphics[width=\linewidth]{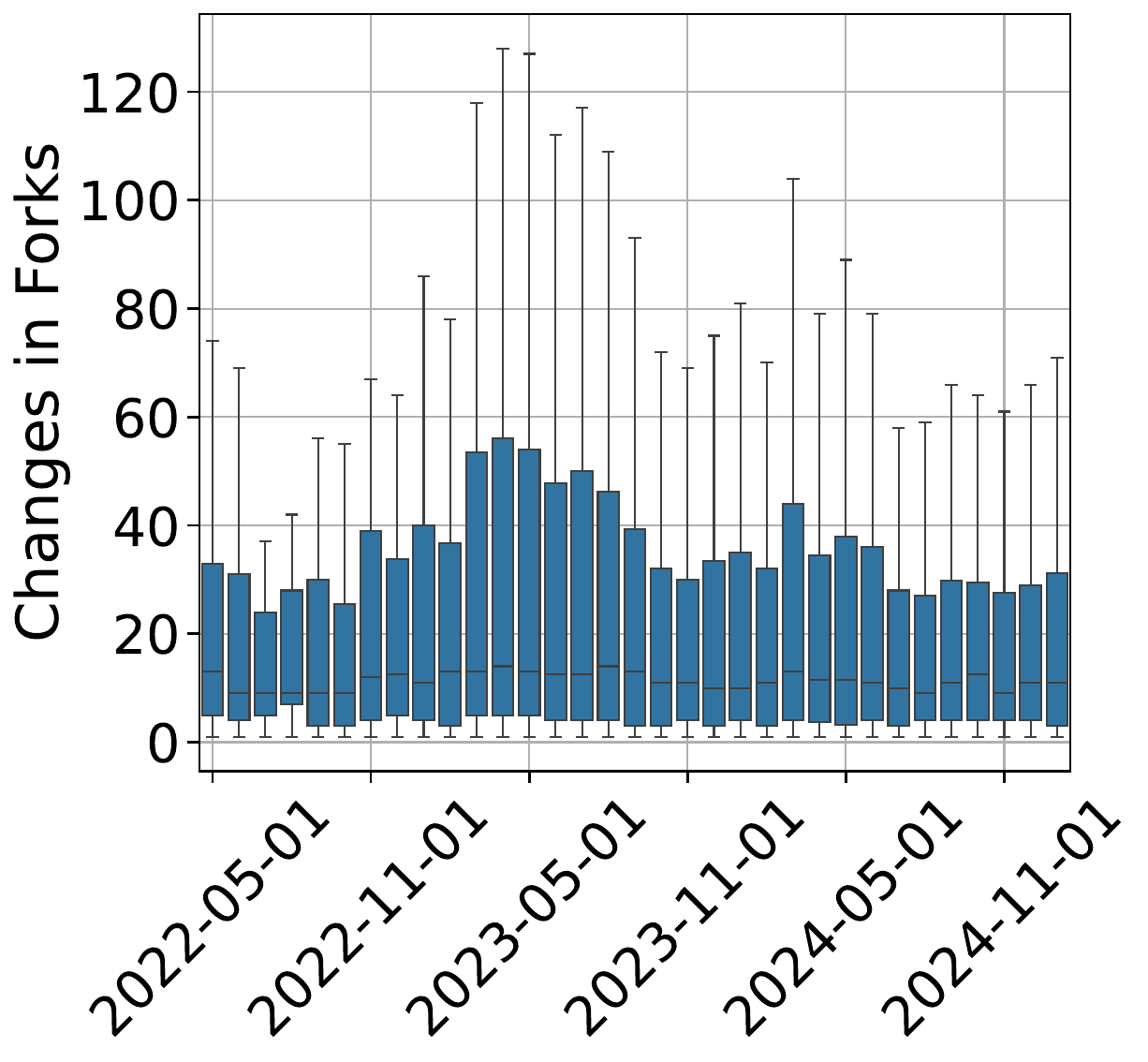}
        \caption{Forks}
        \label{fig:rq3-boxplot-forks}
    \end{subfigure}
    \hfill
    \begin{subfigure}[b]{0.19\textwidth}
        \includegraphics[width=\linewidth]{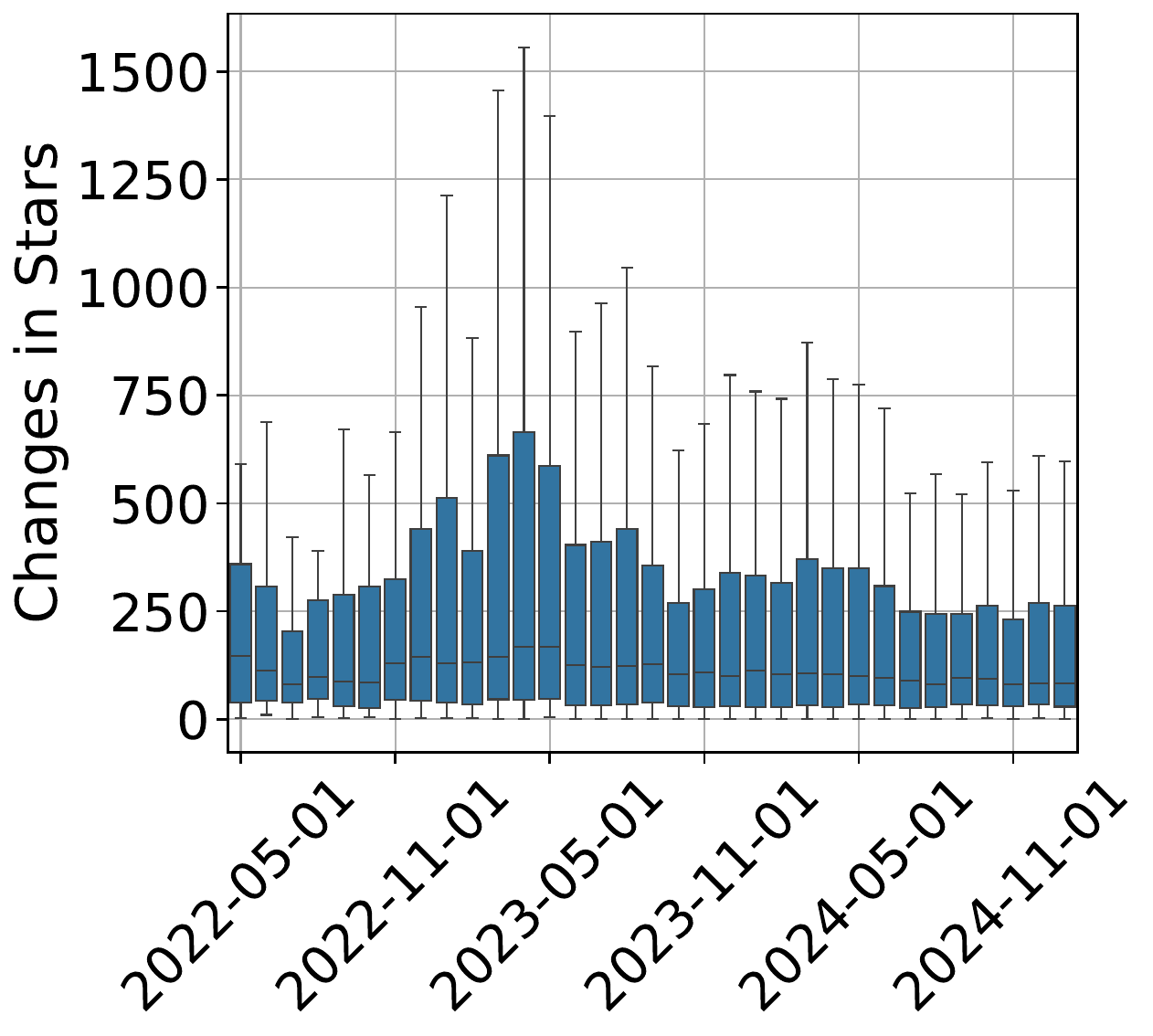}
        \caption{Stars}
        \label{fig:rq3-boxplot-stars}
    \end{subfigure}
    \hfill
    \begin{subfigure}[b]{0.19\textwidth}
        \includegraphics[width=\linewidth]{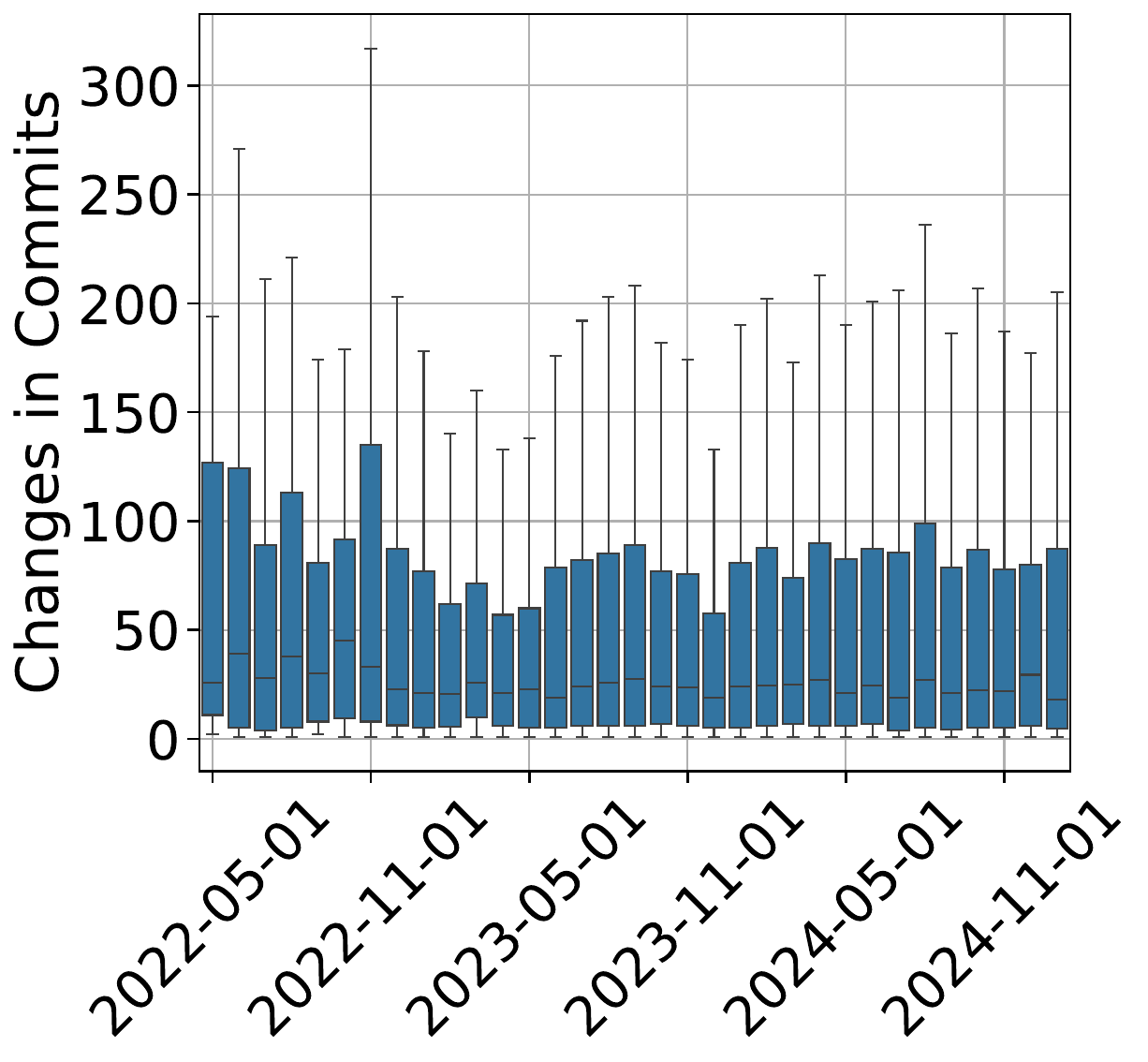}
        \caption{Commits}
        \label{fig:rq3-boxplot-commits}
    \end{subfigure}
    \hfill
    \begin{subfigure}[b]{0.19\textwidth}
        \includegraphics[width=\linewidth]{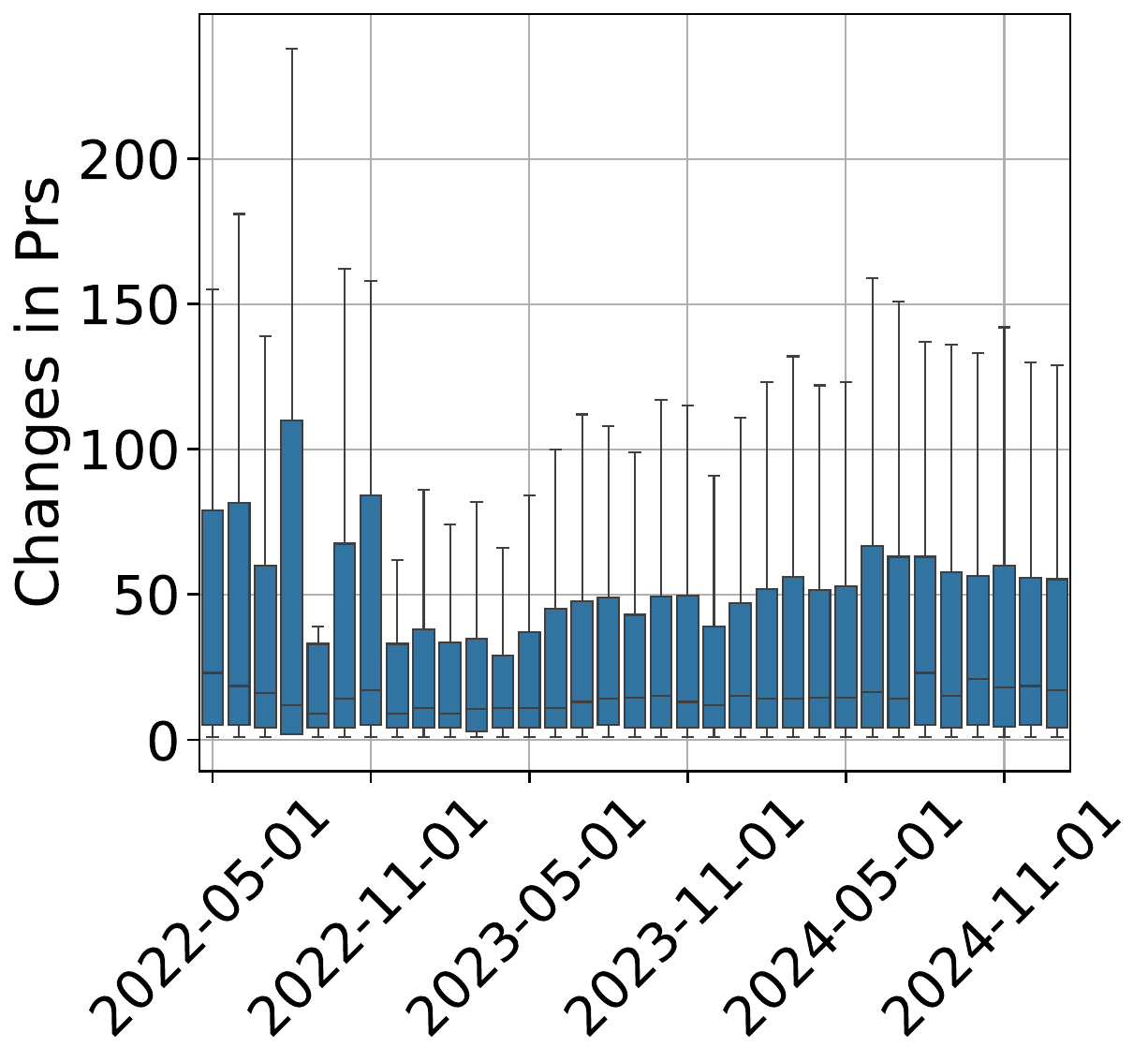}
        \caption{Pull Requests}
        \label{fig:rq3-boxplot-prs}
    \end{subfigure}
    
    \caption{GH-AI projects' metric changes over 2 years since May 8, 2022 (6 months before ChatGPT release) to May 9, 2024 (18 months after ChatGPT release)}
    \label{fig:rq3-plots}
\end{figure*}

\begin{figure*}
    \centering
    \begin{subfigure}[b]{0.19\textwidth}
        \includegraphics[width=\linewidth]{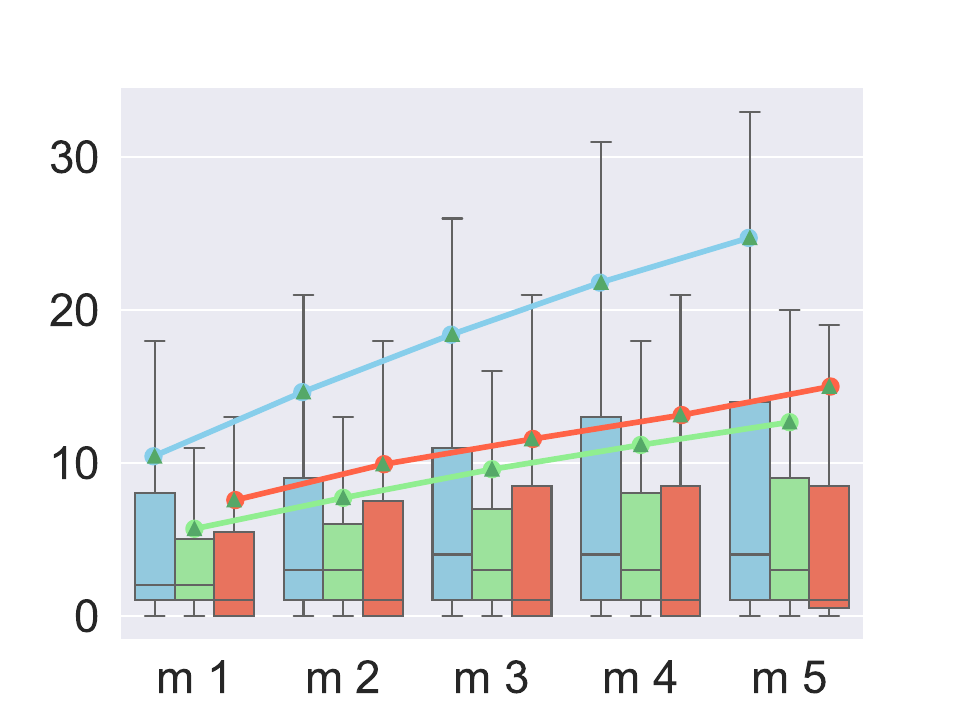}
        \caption{Contributors Growth}
        \label{fig:rq3-submission-contributors}
    \end{subfigure}
    \hfill
    \begin{subfigure}[b]{0.19\textwidth}
        \includegraphics[width=\linewidth]{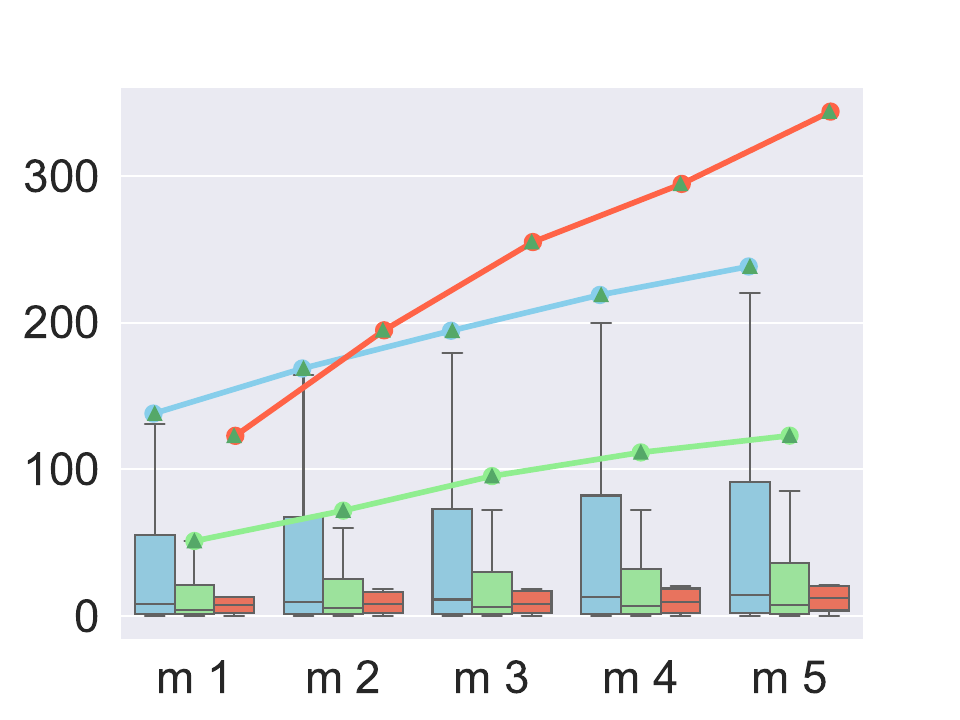}
        \caption{Forks Growth}
        \label{fig:rq3-submission-forks}
    \end{subfigure}
    \hfill
    \begin{subfigure}[b]{0.19\textwidth}
        \includegraphics[width=\linewidth]{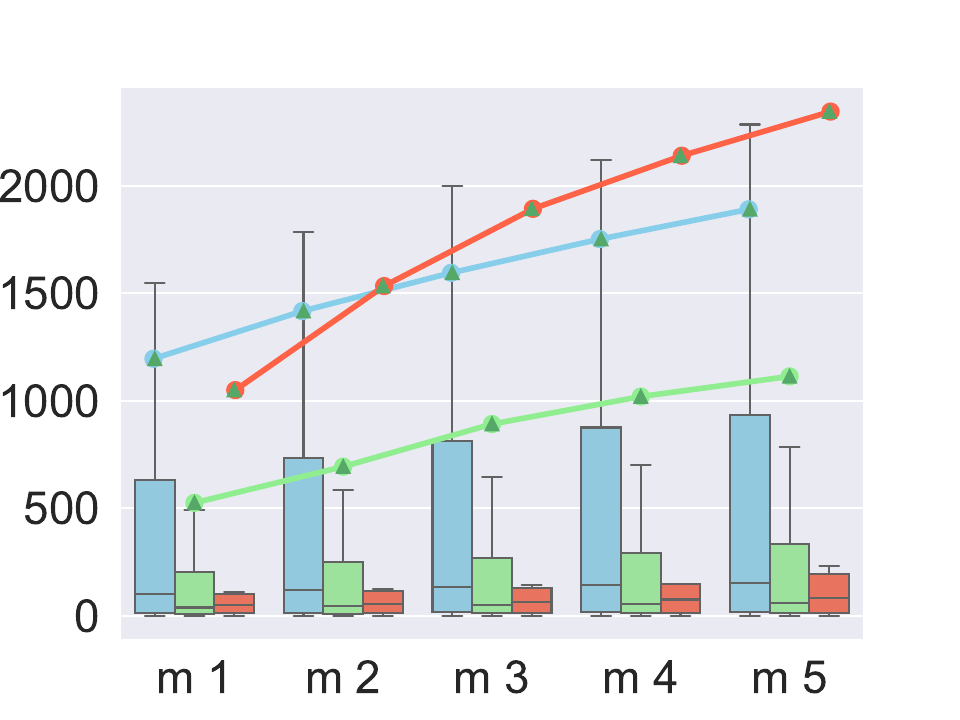}
        \caption{Stars Growth}
        \label{fig:rq3-submission-stars}
    \end{subfigure}
    \hfill
    \begin{subfigure}[b]{0.19\textwidth}
        \includegraphics[width=\linewidth]{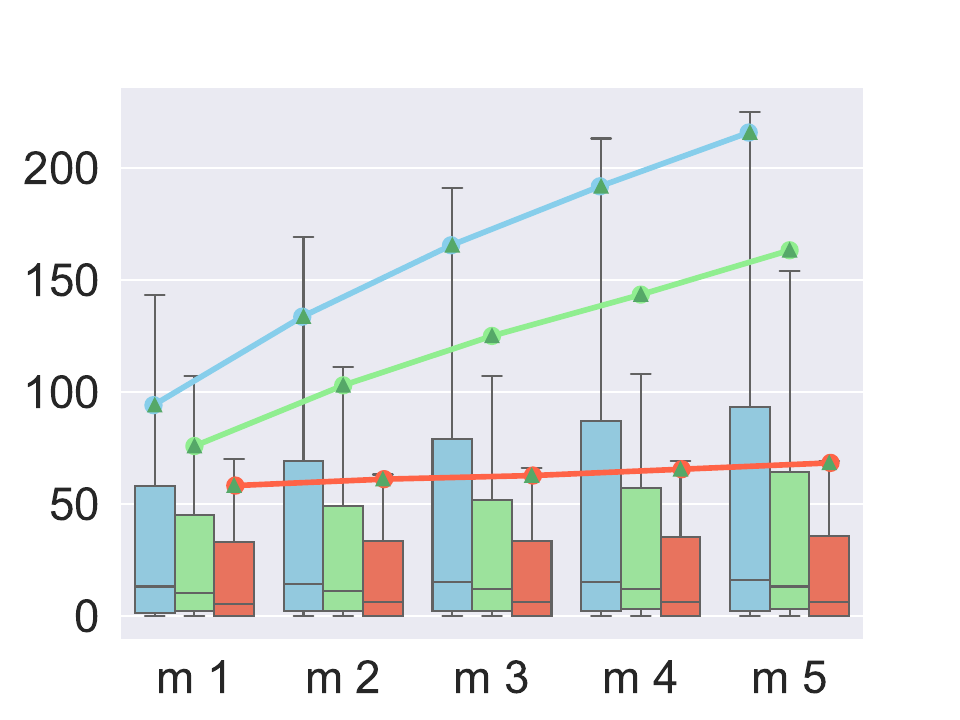}
        \caption{Commits Growth}
        \label{fig:rq3-submission-commits}
    \end{subfigure}
    \hfill
    \begin{subfigure}[b]{0.19\textwidth}
        \includegraphics[width=\linewidth]{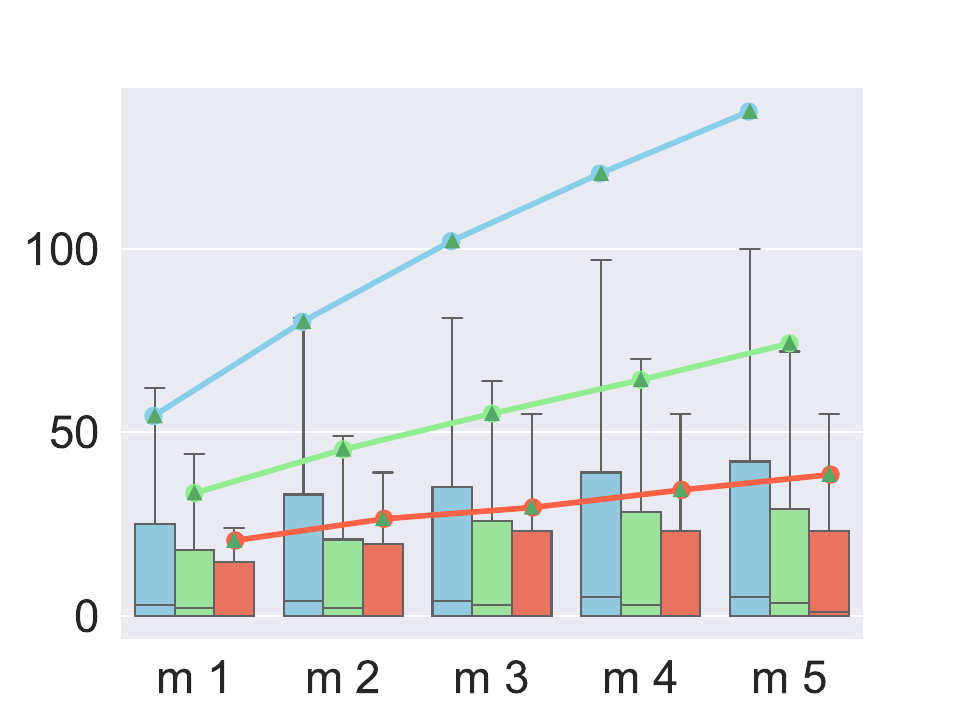}
        \caption{PRs Growth}
        \label{fig:rq3-submission-prs}
    \end{subfigure}
    
    \caption{Growth metrics for GH repository over 5 months (m1--m5) after HN submission (\textcolor{LimeGreen}{green}=positive sentiment, \textcolor{SkyBlue}{blue}=neutral sentiment, \textcolor{RedOrange}{red}=negative sentiment)}
    \label{fig:rq3-submission-growth}
\end{figure*}

\subsubsection{GitHub Metrics Changes}
We observed strong increases in the number of stars and forks metrics of HN GH-AI projects after the release of ChatGPT in late 2023, as shown in Figure~\ref{fig:rq3-plots}. This trend aligns with the findings in RQ1, which show a surge in the number of HN GH-AI stories during the same period. Both commits and contributors exhibit no observable changes, and pull request changes appear to have decreased around late 2022.


However, while the GitHub metric changes gradually decrease over the year after the ChatGPT release, Figure~\ref{fig:hn-trends-gh} indicates a steady number of new HN discussions post-release. Comparing these findings with the results from RQ1, we found that despite a consistent number of new GitHub AI projects, the growth in GitHub metrics, particularly stars and forks, appears to be slowing over time.

\begin{tcolorbox}
Observation 7: GitHub AI projects on Hacker News received an increment of activities on stars and forks shortly after the release of ChatGPT. 
\end{tcolorbox}

\subsubsection{Potential Impact of Hacker News Exposure}
Focusing on the GitHub projects that were posted on Hacker News,
Figure~\ref{fig:rq3-submission-growth} demonstrates the growth of GitHub metrics, including contributors, forks, stars, commits, and pull requests over five months after the Hacker News submission date. They are categorized based on the sentiment of the HN comments in RQ2. Green bars, blue bars, and red bars represent GH AI projects with positive, neutral, and negative comments. 
We can see that GitHub AI projects with positive sentiment in the comments generally have higher average growth (green line) of commits and pull requests after their HN submissions than the neutral and negative sentiment projects.

To determine whether changes in GitHub repository metrics after the Hacker News submission were statistically significant from before the Hacker News submission, we conducted a statistical test comparing the values of each metric one month before (control group) and one month after HN submission (treatment group). 
We included only the GitHub AI repositories that existed for at least six months before their posts appeared on HN to make sure that the changes in the metrics were not affected by the early activities during the creation of the repositories themselves. This resulted in 163 remaining GitHub AI projects for the treatment group. 
Using the non-parametric Wilcoxon signed-rank test~\cite{conoverPracticalNonparametricStatistics1999}, we found statistically significant increases in terms of stars (p-value = 0.001), forks (p-value = 0.004), and contributors (p-value = 0.021). Thus the increased number of stars, forks, and contributors after HN submissions did not happen by chance. 

\begin{tcolorbox}
Observation 8: GitHub AI projects with positive comments on Hacker News showed a higher number of commits and pull requests after HN submission.
\end{tcolorbox}

\textbf{To answer RQ3, we discovered that GitHub AI projects with positive sentiment showed higher activity closely after being posted on Hacker News. This can potentially be affected by their posts on Hacker News. Moreover, most of the analyzed projects had an increase in stars and forks after the release of ChatGPT.}



\section{Implications}
\label{sec:implications}
The observations made from our studies offer a few implications for software engineering researchers and AI developers.

\subsubsection{\textbf{Implications for SE researchers}}This study complements the existing studies~\cite{stoddard2015popularity, Aniche2018, Barik2015} that Hacker News can be used as another important source of information for empirical software engineering research, especially AI development. We found that AI software developers post links to their GitHub projects on Hacker News to attract the attention of IT enthusiasts, and many of them receive positive feedback. Thus, future empirical studies on open-source projects should consider including Hacker News as one of their potential data sources, besides other social platforms like Reddit or well-known and well-studied developer social platforms like Stack Overflow.

\subsubsection{\textbf{Implications for AI developers}}Our study reveals that the Hacker News community is quite positive about AI projects. Thus, advertising AI projects on Hacker News is an effective way to gain attention from the IT community. Our results of the increment in the five GitHub metrics right after posting on Hacker News show that posting the projects on Hacker News may help increase the amount of attention (i.e., stars and forks) and the activities in the projects (i.e., contributors, commits, and pull requests).

\subsubsection{\textbf{Implications for the HN Community}} Our study confirms that there is a community on HN that is connected to developers on GitHub. As shown in RQ3, we see that social media exposure does influence the project activity levels. Hence, especially for projects looking for contributors, HN could be a viable option. The HN community is also shown to be positive and receptive (RQ2). Results from RQ1 also indicate that developers on GitHub are quick to promote their projects on HN.

\section{Related Work}
\label{sec:relatedwork}

\subsection{Topic modeling using Latent Dirichlet Allocation (LDA)}
\label{Ch:Latent_Dirichlet_Allocation}
Topic modeling is a technique used to discover hidden themes or topics within a collection of texts~\cite{Kherwa2019_topicModelling}. Among various topic modeling approaches, LDA \cite{blei2003latent} has become one of the most widely used methods. LDA works by treating each document as a mixture of topics and each topic as a mixture of words. For example, in a collection of technology news articles, one article might be 70 percent about AI development and 30 percent about business implications, while another might be 60 percent about technical implementation and 40 percent about ethical concerns.
A critical step in applying LDA is determining the optimal number of topics. This is typically done by evaluating topic coherence scores across different numbers of topics. Topic coherence measures how semantically similar the words within each topic are, providing a quantitative way to assess topic interpretability~\cite{kapadiaEvaluateTopicModels2022}. 

\subsection{Sentiment Analysis}
\label{sec:sentiment_Analysis}

There are several sentiment analysis techniques proposed and adopted in the software engineering literature. 
These include rule-based sentiment analysis tools~\cite{Lin2018,Zhang2020}, machine learning-based models~\cite{Lin2018,Zhang2020}, pre-trained transformer models~\cite{Zhang2020,Zhang2025}, and prompted LLMs~\cite{Zhang2024,Zhang2025}.
Lin et al.~\cite{Lin2018} utilized SentiStrength, NLTK, Standford CoreNLP, SentiStrength-SE, and Standford CoreNLP SO on three software engineering datasets. Building on~\cite{Lin2018}, Zhang et al.~\cite{Zhang2020} further explored pre-trained transformer models, including BERT, RoBERTa, XLNet, and ALBERT, for sentiment analysis across multiple software engineering datasets. More recently, Zhang et al.~\cite{Zhang2024} investigated the general capability of LLMs for sentiment analysis. 
In the context of software engineering, Zhang et al.~\cite{Zhang2025} conducted a study investigating the prompt sensitivity of bigger language models (BLMs), such as Llama-2-13b-chat-hf, Vicuna-13b-v1.5, and WizardLM-13B-V1.2, as well as evaluating their performance against fine-tuned smaller language models (SLMs), including BERT, RoBERTa, XLNet, and ALBERT, across five software engineering datasets.
These studies prompted us to explore multiple approaches for sentiment analysis on Hacker News stories and comments, including fine-tuning pre-trained transformer models and LLM.




\subsection{Hacker News Studies}
 
There are a few studies that focus on Hacker News.
Stoddard et al.~\cite{stoddard2015popularity} investigate the relationship between an online article's popularity and its inherent quality on social news aggregators. The study proposes a method using Poisson regression to estimate an article's quality. By analyzing time-series voting data from Reddit and Hacker News, the study finds a strong correlation between observed popularity and the estimated intrinsic quality, particularly among articles receiving a reasonable amount of attention. 
Barik et al.~\cite{Barik2015} propose a novel approach to enhance grounded theory research in software engineering by leveraging Hacker News data. The authors suggest automatically extracting knowledge from online discussions to replicate and validate existing findings. Their proof-of-concept study on static analysis tool adoption using Hacker News demonstrates the potential to confirm existing themes, discover new ones like security and ego, and ultimately automate parts of the analysis process. 
Another study by Aniche et al.~\cite{anicheHowModernNews2018} explores how software developers utilize Reddit and Hacker News to share and shape knowledge within their communities.

\subsection{Studies Related to Other Social Media for Software Development}



\section{Threats to Validity}
\textit{Internal Validity:~}The keywords used for collecting Hacker News AI stories may not cover all possible AI-related stories.
Moreover, we relied on automated sentiment analysis, which may create some false positives or false negatives. We mitigated this by selecting the best-performing technique from a comparison of several state-of-the-art sentiment analysis techniques.

\textit{External validity:~}This study focuses on Hacker News. The findings may not be generalized to other social news aggregators such as Reddit. Moreover, the analyzed GitHub projects are OSS AI-related projects, so the results may not be applicable to other GitHub projects or commercial projects.
\label{sec:threats}

\section{Conclusion}
\label{sec:conclusion}

We analyze 2,195 Hacker News submissions and their associated comment threads over a two-year period. The findings indicate that AI developers actively promoted their GitHub repositories on Hacker News. An investigation of the reaction using sentiment analysis shows that Hacker News stories related to GitHub AI projects frequently receive positive engagement from the community. Subsequent analysis of the corresponding GitHub repositories reveals a significant increase in development activity, measured through metrics such as commits, pull requests, forks, stars, and the number of contributors, following promotion on the platform.

\bibliographystyle{IEEEtran}
\bibliography{references}

\end{document}